\documentclass{jfm}
\usepackage{natbib}
\usepackage{upmath}
\usepackage[british]{babel}
\usepackage{amsmath,bm}
\usepackage{amssymb}
\usepackage{amsbsy}
\usepackage{amscd}
\usepackage{amstext}
\usepackage{tabularx}
\usepackage{float}
\usepackage{makeidx}
\usepackage{amsmath}
\usepackage{subfigure}
\usepackage{afterpage}
\usepackage[T1]{fontenc}
\usepackage[latin1]{inputenc}
\usepackage{multirow}
\usepackage{color}
\usepackage{graphicx} % not used later, figures must be supplied as hardcopies, use \vspace
\usepackage{graphics,epsfig}
\usepackage{amsfonts}
\usepackage{psfrag}

\ifCUPmtlplainloaded \else
  \checkfont{eurm10}
  \iffontfound
    \IfFileExists{upmath.sty}
      {\typeout{^^JFound AMS Euler Roman fonts on the system,
                   using the 'upmath' package.^^J}%
       \usepackage{upmath}}
      {\typeout{^^JFound AMS Euler Roman fonts on the system, but you
                   dont seem to have the}%
       \typeout{'upmath' package installed. JFM.cls can take advantage
                 of these fonts,^^Jif you use 'upmath' package.^^J}%
      }
  \else
  \fi
\fi

\ifCUPmtlplainloaded \else
  \checkfont{msam10}
  \iffontfound
    \IfFileExists{amssymb.sty}
      {\typeout{^^JFound AMS Symbol fonts on the system, using the
                'amssymb' package.^^J}%
       \usepackage{amssymb}%

      }{}
  \fi
\fi

\ifCUPmtlplainloaded \else
  \IfFileExists{amsbsy.sty}
    {\typeout{^^JFound the 'amsbsy' package on the system, using it.^^J}%
     \usepackage{amsbsy}}
    {\providecommand\boldsymbol[1]{\mbox{\boldmath $##1$}}}
\fi

\newsavebox{\astrutbox}
\sbox{\astrutbox}{\rule[-5pt]{0pt}{20pt}}

\newcommand\bu{\boldsymbol{u}}
\newcommand\bc{\boldsymbol{c}}
\newcommand\bgamma{\boldsymbol{\gamma}}

\newcommand{\ri}{\mathop{\rm i}\nolimits}
\newcommand{\re}{\mathop{\rm e}\nolimits}

\title[On the large-$Wi$ scaling laws in viscoelastic pipe flows]
{On the large-Weissenberg-number scaling laws in viscoelastic pipe flows}

\author[D. Wan, M. Dong, M. Zhang]%
{
Dongdong Wan$^1$, Ming Dong$^2$ \and Mengqi Zhang$^{1}$
}

\affiliation{
$^1$Department of Mechanical Engineering, National University of Singapore, 9 Engineering Drive 1, 117575 Singapore \\
$^2$State Key Laboratory of Nonlinear Mechanics, Institute of Mechanics, Chinese Academy of Sciences, Beijing 100190, China
  \\[\affilskip]
}

\date{xx}

\graphicspath{{figures/}}
\newlength\savewidth

\begin{document}
\maketitle

\begin{abstract}

This work explains a scaling law of the first Landau coefficient of the derived Ginzburg-Landau equation (GLE) in the weakly nonlinear analysis of axisymmetric viscoelastic pipe flows in the large-Weissenberg-number ($Wi$) limit, recently reported in Wan \textit{et al.} \textit{J. Fluid Mech.} (2021), vol. 929, A16. Using an asymptotic method, we derive a reduced system, which captures the characteristics of the linear centre-mode instability near the critical condition in the large-$Wi$ limit. Based on the reduced system we then conduct a weakly nonlinear analysis using a multiple-scale expansion method, which readily explains the aforementioned scaling law of the Landau coefficient and some other scaling laws. Particularly, the equilibrium amplitude of disturbance near linear critical conditions is found to scale as $Wi^{-1/2}$, which may be of interest to experimentalists. The current analysis reduces the numbers of parameters and unknowns and exemplifies an approach to studying the viscoelastic flow at large $Wi$, which could shed new light on the understanding of its nonlinear dynamics.

\end{abstract}

\section{Introduction}

Significant skin drag reduction occurs when few parts per million of polymers are added to turbulent Newtonian flows \citep{Toms1949,Virk1975}. This drag-reduction mechanism has been successfully applied in the Trans-Alaska pipeline project \citep{Burger1980}. Drag reduction via polymer ejection around the marine vehicle hull has also been reported: the marine vehicle's speed can increase by up to 15\% due to the drag-reducing effect of polymers \citep{national1997}. Because of its tremendous economic potential, continuous research effort has been devoted to studying this phenomenon \citep{White2008,Graham2014Drag,Datta2021,Steinberg2021,Sanchez2022Understanding}.

In this subject, one of the most important but unsolved problems concerns how a viscoelastic flow transitions from a laminar state to turbulence.
From a general perspective, the Newtonian pipe flow is linearly stable even at a large Reynolds number (or $Re$) according to numerical evidence \citep{Davey1969,Meseguer2003Linearized}, but it transitions to turbulence at a low $Re$ \citep{Avila2011}, suggestive of a nonlinear subcritical transition route at play. Polymer additives enrich the dynamics of the Newtonian flow and provide new possibilities for flow transition. Especially, polymers render the flow elastic.
In the current literature on viscoelastic flows, two transition routes are actively researched in a large portion of Reynolds number-Weissenberg number ($Re-Wi$) space. One of them is the elastically-modified wall mode mediating the classical Newtonian subcritical transition route when flow elasticity is relatively weak \citep{Shekar2019Critical-Layer,Shekar2021} and the other is a centre-mode instability causing a supercritical transition to elasto-inertial turbulence (EIT) when the elastic effect is strong \citep{Garg2018Viscoelastic}. The present work studies scaling laws related to the centre mode in viscoelastic pipe flows.

\cite{Garg2018Viscoelastic} first reported the centre-mode instability in viscoelastic pipe flows. The unstable mode was found based on an Oldroyd-B constitutive model with a phase speed close to the maximum velocity of the laminar flow. It is believed that this centre-mode instability can lead the flow to EIT, which is plausibly related to the maximum drag reduction (MDR) \citep{Samanta2013Elasto-inertial}. At the onset of this instability, the $Re$ is significantly smaller than the typical nonlinear critical $Re$ for Newtonian turbulence in pipe and $Wi$ is larger than order 1. Scaling laws of the linear critical Reynolds number $Re_c$ and the linear critical wavenumber $\alpha_c$ have been derived in \cite{Garg2018Viscoelastic}, namely $Re_c\sim [E(1-\beta)]^{-3/2}$ and $\alpha_c\sim [E(1-\beta)]^{-1/2}$ for $E(1-\beta)\ll1$ (see also \cite{Chaudhary2021Linear}; here $E\equiv Wi/Re$ is the elasticity number and $\beta$ the solvent-to-solution viscosity ratio). The authors explained these scaling laws using a regular perturbation technique. Instead, by means of an asymptotic technique, \cite{Dong2022Asymptotic} analysed the same flow in the large-$Re$ limit to explain these scaling laws. The authors found a three-layered structure of the centre-mode instability in both a long-wavelength regime and a short-wavelength regime. These are linear results. Later, \cite{Page2020Exact} calculated the exact travelling wave solutions in two-dimensional viscoelastic channel flows using arclength continuation starting from the unstable centre mode (see also \cite{Khalid2021Centre}). They established the subcriticality of the flow transitioning to EIT, in addition to the supercritical route.
{More recently, \cite{Buza2022Finite,Buza2021} and \cite{Morozov2022Coherent} have further calculated the finite-amplitude travelling wave solutions in the inertialess limit, extending the linear instability found by \cite{Khalid2021}. These solutions are believed to be the underlying mechanism for the transition to elastic turbulence observed in experiments \citep{Jha2020Universal}. } These nonlinear travelling wave solutions, being saddle points in a certain state space, can act as the building blocks of the spatially and temporally chaotic turbulent flow (see reviews by \cite{Kerswell2005,Eckhardt2006,Graham2021Exact}).

The previous works thus implied possibilities of both subcritical and supercritical transitions in viscoelastic flows. These two bifurcation routes have also been observed in experiments. \cite{Samanta2013Elasto-inertial} first studied experimentally the EIT phenomenon in a viscoelastic pipe flow. When the polymer concentration is low, they found a clear hysteresis loop when changing $Re$, indicating the existence of a subcritical transition mechanism, whereas, when the polymer concentration is high, a non-hysteresis behaviour was observed, implying a supercritical transition route, even though the authors warned that the polymeric flows may be sensitive to disturbances.
\cite{Chandra2020Early} experimentally studied the polymeric flow in microtubes and found a cross-over of the transition route from subcritical to supercritical as the polymeric effect strengthens.
Recent experiments by \cite{Choueiri2021} on the transition to EIT in viscoelastic pipe flows also confirmed the chevron shaped streaks consistent with the linear stability theory (centre mode) and revealed a secondary instability related to a wall mode at subcritical $Re$.

As background disturbance inherently exists in experiments, determining the bifurcation type of a sensitive flow is always difficult.
On the other hand, it is enlightening to systematically study the bifurcation type in viscoelastic pipe flows from the governing equations (of course, model defects are also inevitable, dealing with which is however beyond the scope of this work; here we employ the Oldroyd-B model, which has been used previously in \cite{Garg2018Viscoelastic}, \cite{Chaudhary2021Linear}, etc.).
Weakly nonlinear stability analyses have been traditionally applied to study the flow bifurcation. The theory was originally proposed by \cite{Landau1944} and developed later by many researchers in the hydrodynamic stability community \citep{Stuart1960Non-linear,Reynolds1967Finite,Herbert1983a,Fujimura1989Equivalence}. In the context of viscoelastic flows, Morozov \& van Saarloos and their co-workers have applied the weakly nonlinear stability analysis to both pipe and channel Poiseuille flows of Oldroyd-B fluids and Upper Convected Maxwell (UCM) fluids in the inertialess limit \citep{Meulenbroek2003Intrinsic,Meulenbroek2004Weakly,Morozov2019Subcritical}, demonstrating a generic nonlinear subcritical instability in these flows. Regarding the bifurcation in EIT, two recent studies, \cite{Wan2021Subcritical} and \cite{Buza2021}, have performed the weakly nonlinear stability analyses of viscoelastic finite-$Re$ pipe and channel flows, respectively.
Using a multiple-scale expansion method, \cite{Buza2021} explored the subcriticality of viscoelastic channel flows to lower $Wi$ required by the linear instability to reveal the purely elastic nature of the instability (see also \cite{Khalid2021}). \cite{Wan2021Subcritical} adopted the same method to determine the bifurcation type of viscoelastic pipe flows in a large parameter space. They derived a Ginzburg-Landau equation (GLE) from the Navier-Stokes equations and the polymer constitutive  equations around linear critical conditions. Their theoretical results indicate that when the viscosity ratio $\beta$ is large, the viscoelastic pipe flow experiences a subcritical transition, whereas, when it is small, the flow will transition supercritically from the laminar state, consistent with the experimental observations summarised above. Besides, they found a scaling law of the third-order Landau coefficient $a_3$ in GLE with $Wi$: the value of $a_3$ scales with $1/Wi$ at a fixed $\beta$ when $Wi$ is sufficiently large (see $a_3$ in Eq. \eqref{eq:GL-tau2 scale} to follow).

Research on scaling law has deepened our understanding of the flow transition. Studies exist on the scaling law of the disturbance amplitude threshold (beyond which a transition initiates in the subcritical regime) in Newtonian flows. An asymptotic analysis of the Newtonian channel flow by \cite{Chapman2002Subcritical} showed that the transitional threshold amplitude scales with $Re^{-3/2}$, which was later experimentally verified by \cite{Philip2007Scaling} (with $Re^{-1.53}$ for $1000<Re<2000$). For $2000<Re<5000$, \cite{Lemoult2012Experimental} found the scaling to be $Re^{-1}$, consistent with the theoretical prediction by \cite{Waleffe2005Transition}. Similarly, in experiments on Newtonian pipe flows \cite{Hof2003Scaling} uncovered a scaling of $Re^{-1}$, which also agrees with theoretical results. 
In viscoelastic flows, \cite{Jovanovic2010Transient} derived scaling laws of the non-modal transient growth in inertialess plane Couette and Poiseuille flows. This non-modal mechanism is believed to underlie the elastic instability in perturbed channel flows \citep[e.g.][]{Shnapp2021}.
\cite{Morozov2007} derived a scaling of $Wi^{-2}$ for the amplitude threshold in the flow transition at low $Re$. The derivation was based on a nonlinear flow instability criterion proposed in \cite{Pakdel1996Elastic} for elastic instabilities. Inspired by these works, we in this paper study and explain the scaling laws in the nonlinear regime of viscoelastic pipe flows (first found by \cite{Wan2021Subcritical}).

In the following, we will derive a reduced nonlinear system from the original governing equations at asymptotically large $Wi$ (thus, the reduced system does not depend on $Wi$ \textit{explicitly}) using an asymptotic method (Sec. 2), extending the scaling analysis in \cite{Garg2018Viscoelastic} and \cite{Chaudhary2021Linear}. We show that the numerical results of \cite{Wan2021Subcritical} indeed converge asymptotically to those of the reduced system when $Wi$ increases (Sec. 3). Then, a multiple-scale expansion of the reduced system is conducted to bring out the scaling law of $a_3$ with $Wi^{-1}$ after $Wi$ is re-introduced back to the system (and some other laws, see Eq. \ref{eq:full-redu-trans}); in particular, a scaling law of the equilibrium amplitude of disturbance is derived from the GLE. Sec. 4 concludes the paper with some discussions.

\section{Problem formulation}\label{problemformulation}

\subsection{Governing equations and parameters}

We investigate the hydrodynamic stability of incompressible pipe Poiseuille flows based on the Oldroyd-B fluid model \citep{Bird1987Dynamics}. The cylindrical coordinate system is used, with $r$, $\theta$ and $z$ denoting the radial, azimuthal and axial directions, respectively. The pipe radius $R$ and the centreline velocity $U_c$ are chosen to be the characteristic length and velocity scales to normalise the system. The non-dimensional perturbation equations are \citep{Wan2021Subcritical} 
\begin{subequations}\label{eq:disturbance}
	\begin{equation}
	\boldsymbol{\nabla}  \cdot \bu = 0, \ \ \ \ \ \ \ \partial_t \bu + \bu \cdot \boldsymbol{\nabla} \boldsymbol{U} + \boldsymbol{U} \cdot \boldsymbol{\nabla} \bu  + \boldsymbol{N}_{\bu} = - \boldsymbol{\nabla} p + \frac{\beta}{Re} \nabla^{2} \bu + \frac{1-\beta}{ReWi} \boldsymbol{\nabla} \cdot \bc,
	\end{equation}
	\begin{equation}
	{\partial_t} \bc + \bu \cdot \boldsymbol{\nabla} \boldsymbol{C} - \bc \cdot \boldsymbol{\nabla} \boldsymbol{U} - (\boldsymbol{\nabla} \bu)^{T} \cdot \boldsymbol{C} + \boldsymbol{U} \cdot \boldsymbol{\nabla} \bc - \boldsymbol{C} \cdot \boldsymbol{\nabla} \bu - (\boldsymbol{\nabla} \boldsymbol{U})^{T} \cdot \bc  + \boldsymbol{N}_{\bc} = - \frac{\bc}{Wi},
	\end{equation}
\end{subequations}
where $\bu = (u_{r}, u_{\theta}, u_{z})^T$ is the perturbation velocity vector, $\bc=(c_{rr}, c_{r\theta}, c_{rz}, c_{\theta\theta}, c_{\theta z}, c_{zz})^T$ the  conformation tensor and $p$ the  pressure. $\boldsymbol{U}=(0,0,U_z)$ and $\boldsymbol{C}=(1,0,WiU_z',1,0,1+2Wi^2U_z'^2)$ are the corresponding laminar base states, where $U_z=1-r^2$ and prime $'$ denotes differentiation with respect to $r$. The nonlinear terms are $\boldsymbol{N}_{\bu} =  \bu \cdot \boldsymbol{\nabla} \bu$ and $\boldsymbol{N}_{\bc} =  \bu \cdot \boldsymbol{\nabla} \bc - \bc \cdot \boldsymbol{\nabla} \bu - (\boldsymbol{\nabla}\bu)^{T} \cdot \bc$. The controlling parameters include the viscosity ratio $\beta=\frac{\nu_s}{\nu_s+\nu_p}$ (where $\nu_s$ is solvent viscosity and $\nu_p$ polymer viscosity), the Reynolds number $Re=\frac{U_c R}{\nu_s+\nu_p}$ and the Weissenberg number $Wi=\frac{\lambda U_c}{R}$ ($\lambda$: polymer relaxation time). We define the elasticity number as $E\equiv Wi/Re=\frac{\lambda (\nu_s+ \nu_p)}{R^2}$, characterising the elastic effects of polymers. The no-slip boundary condition $\bu(r=1)=\boldsymbol{0}$ is applied at the pipe wall. Note that for the conformation tensor, it is not necessary to specify its boundary condition, because the linear operator in equation (\ref{eq:disturbance}$b$) does not contain $r$-derivative terms (like $\partial/\partial_r$) of $\bc$ (although such terms do exist in the nonlinear term $\boldsymbol{N}_{\bc}$, which, in a weakly nonlinear framework, is constructed from the linear eigenvectors, as shown below in equation \eqref{eq:nonlinear_N2_N3}).
Following previous works \citep{Garg2018Viscoelastic,Chaudhary2021Linear,Zhang2021}, we are interested in the axisymmetric disturbance because only this mode is found to be linearly unstable. Therefore, the symmetric conditions are imposed at the pipe centre, $u_r(r=0)= u'_z(r=0)=0$.

The equation system \eqref{eq:disturbance} will hereafter be referred to as the original equation system and we use subscript "F" to mark it. Introducing $\boldsymbol{\gamma}_{\text{F}}=(u_{r\text{F}},u_{z\text{F}},p_{\text{F}},c_{rr\text{F}},c_{rz\text{F}},c_{\theta\theta\text{F}},c_{zz\text{F}})^T$,
the equation system can be recast to a compact form of
\begin{equation}\label{eq:original_compact_nonlinear}
(\boldsymbol{M}_{\text{F}} {\partial_t}  -\boldsymbol{L}_{\text{F}} )\bgamma_{\text{F}}= \boldsymbol{N}_{\text{F}},
\end{equation}
where the weight matrix $\boldsymbol{M}_{\text{F}}$, the linear operator $\boldsymbol{L}_{\text{F}}$ and the nonlinear operator $\boldsymbol{N}_{\text{F}}$ can readily be derived from \eqref{eq:disturbance}. The linear mode is governed by the homogeneous system
\begin{equation}\label{eq:original_compact_linear}
(\boldsymbol{M}_{\text{F}} {\partial_t}  -\boldsymbol{L}_{\text{F}} )\bgamma_{\text{F}}= \boldsymbol{0},
\end{equation}
which admits the normal mode wavelike solution
\begin{equation}\label{eq:original_wavelike}
{\bgamma}_{\text{F}}=\tilde {\bgamma}_{\text{F}}(r)\re^{\ri\alpha (z-c t)} + c.c.
\end{equation}
Here $\tilde {\bgamma}_{\text{F}}$ is the linear eigenfunction, $\ri$ is the imaginary unit, $\alpha$ is the axial wavenumber, $c=c_r + \ri c_i$ is the complex phase speed ($\omega=\alpha c$ is the complex frequency) with $\alpha c_i$ (or $\omega_i$) representing the linear growth rate, and $c.c.$ represents the complex conjugate of its preceding term.
By substituting \eqref{eq:original_wavelike} into \eqref{eq:original_compact_linear}, one obtains the following linear eigenvalue problem for the flow
\begin{equation}\label{eq:original_compact_linear_eigenvalue}
\left(c \widetilde{\boldsymbol{M}}_{\text{F}}-\widetilde{\boldsymbol{L}}_{\text{F}}\right)\tilde{\bgamma}_{\text{F}}=\boldsymbol{0},
\end{equation}
where $\widetilde{\boldsymbol{M}}_{\text{F}}$ and $\widetilde{\boldsymbol{L}}_{\text{F}}$ can be derived from $\boldsymbol{M}_{\text{F}}$ and $\boldsymbol{L}_{\text{F}}$ respectively by replacing $\partial_z$ with $\ri \alpha$.

\subsection{Asymptotic analysis}\label{sec:asymptotic_analysis}

\begin{figure}
	\centering
	\includegraphics[width=0.5\textwidth,trim= 0 0 0 0,clip]{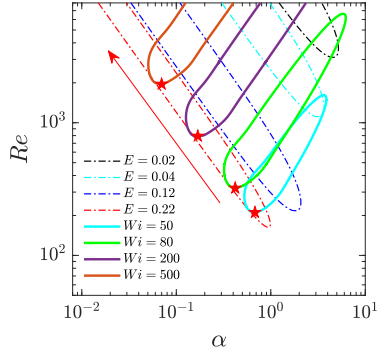}
	\caption{{Neutral curves in the $\alpha$-$Re$ plane at viscosity ratio $\beta=0.65$ for viscoelastic pipe flows based on the Oldroyd-B fluid model. The solid curves are computed for fixed $Wi$, with the linear critical conditions ($Re_c$ and $\alpha_c$) marked by the red stars; following the red arrow direction $\alpha_c$ decreases and is smaller than one, implying the long-wavelength nature of the centre mode. The dot-dashed curves are obtained for fixed $E$ (as shown in figure 18 of \cite{Chaudhary2021Linear}). Adapted from figure 3$(c)$ in \cite{Wan2021Subcritical}.}}
	\label{Fig:NC_beta65}
\end{figure}

Now, we conduct an asymptotic analysis of the centre-mode instability by assuming $Wi\gg 1$ and $E=O(1)$ ($Re \gg 1$). The polymer concentration is taken to be of an intermediate level, i.e., the viscosity ratio $(\beta,1-\beta)=O(1)$.
Solving the linear eigenvalue problem \eqref{eq:original_compact_linear_eigenvalue}, we can obtain the neutral curves in the $\alpha$--$Re$ plane. Figure \ref{Fig:NC_beta65} shows an example for $\beta=0.65$, which is adapted from figure 3$(c)$ of \cite{Wan2021Subcritical}.
For a given $Wi$, the enclosed region by the solid line represents the unstable zone, and its onset at the lowest $Re$, marked by the red star, represents the linear critical condition. Increase of $Wi$ leads to a higher critical Reynolds number $Re_c$ and a lower critical wavenumber $\alpha_c$, as shown by the red arrow.
The control parameters of our interest pertain to these linear critical conditions. From the figure, one can understand that the disturbances to be studied (especially for high $Wi$) are of small $\alpha$, indicating a long-wavelength nature.
Moreover, under these linear critical conditions, the elasticity number $E$ is of $O(1)$ as illustrated in figure 11$(e)$ of \cite{Wan2021Subcritical}. In the figure, we also show the neutral curves for fixed $E$ where the linear critical conditions at low $E$, as traced by \cite{Chaudhary2021Linear}, are related to the short-wavelength regime and will not be considered here.

In our asymptotic analysis, we assume $Wi^{-1} \ll \alpha$, which is consistent with the linear critical conditions in figure \ref{Fig:NC_beta65}. Thus, a small parameter, $\sigma=(\alpha Wi)^{-1} \ll 1$, can be introduced. The complex phase speed is then expanded as
\begin{equation}\label{eq:expansion_c}
c=1+\sigma c_1+\cdots,
\end{equation}
where $c_1$ is the phase speed correction to be solved for. From the balance of the leading-order terms in the linear system \eqref{eq:original_compact_linear_eigenvalue}, we obtain that in the bulk region where $r=O(1)$, $(\tilde u_{r\text{F}},\tilde p_{\text{F}},\tilde c_{rr\text{F}},\tilde c_{rz\text{F}},\tilde c_{zz\text{F}}) \sim (\alpha,1,\sigma^{-1} ,\alpha^{-1}\sigma^{-2}, \alpha^{-2}\sigma^{-2})\tilde u_{z\text{F}}$. For example, from the continuity equation, we know that $\tilde{u}'_{r\text{F}} \sim\ri \alpha \tilde{u}_{z\text{F}}$; thus, we have $\tilde{u}_{r\text{F}} \sim \alpha \tilde{u}_{z\text{F}}$. The leading-order perturbation field can then be rescaled as
\begin{equation}\label{eq:rescaling}
(u_{r\text{F}},u_{z\text{F}},p_{\text{F}},c_{rr\text{F}},c_{rz\text{F}},c_{zz\text{F}}) = (\alpha  u_{r},  u_{z}, p,\sigma^{-1}  c_{rr}, \alpha^{-1}\sigma^{-2}  c_{rz}, \alpha^{-2}\sigma^{-2}  c_{zz})+ \cdots.
\end{equation}
The long-wavelength nature of the instability determines that the radial momentum equation, to the leading order, reduces to $0=-p'_\text{F}$; so $c_{\theta\theta\text{F}}$ does not appear in the leading-order equation system. Because $c$ is close to unity, we introduce the relation
\begin{equation}\label{eq:t_z_tau}
{\partial_t}=-{\partial_z} + \sigma {\partial_\tau},
\end{equation}
where $\tau$ is a time scale related to $c_1$, recalling the expansion of $c$ in \eqref{eq:expansion_c}. Noting that in the long-wavelength limit, both $t$, $\tau$ and $z$ are of $O(\alpha^{-1})$, we introduce
\begin{equation}\label{eq:t_z_tau_bar}
(\bar t, \bar \tau, \bar z)=\alpha (t, \tau, z)=O(1).
\end{equation}

With the above assumptions, applying \eqref{eq:rescaling} in \eqref{eq:original_compact_nonlinear} and neglecting the $O(Wi^{-1})$ terms, we obtain the following asymptotic equations (referred to as the reduced system)
\begin{subequations}\label{eq:asymptotic}
	\begin{equation}\tag{\ref{eq:asymptotic}$a,b$}
	0= u_{r}'+ u_{r}/r + \partial_{\bar z} u_{z}, \ \ \ \ \ \ \ \  \ \ \ 0=- p',
	\end{equation}
	\begin{equation}\tag{\ref{eq:asymptotic}$c$}
	\sigma \partial_{\bar \tau} u_{z}=2r u_{r} + r^2 \partial_{\bar z} u_{z} + \sigma\beta E(u_{z}''+{u_{z}'}/{r}) - \partial_{\bar z} p + (1-\beta)E (c_{rz}'+{c_{rz}}/{r} + \partial_{\bar z}c_{zz}) - N_{u_{z}},
	\end{equation}
	\begin{equation}\tag{\ref{eq:asymptotic}$d$}
	\sigma \partial_{\bar \tau} c_{rr} = 2\sigma u_{r}' - 4r\partial_{\bar z}u_{r} + r^2 \partial_{\bar z} c_{rr}-\sigma c_{rr} - N_{c_{rr}},
	\end{equation}
	\begin{equation}\tag{\ref{eq:asymptotic}$e$}
	\sigma \partial_{\bar \tau} c_{rz} =2\sigma u_r - 2\sigma r u_{r}' + 8 r^2\partial_{\bar z} u_{r} +\sigma^2 u_{z}' - 2 \sigma r \partial_{\bar z} u_{z} - 2\sigma r c_{rr} + r^2 \partial_{\bar z} c_{rz}-\sigma c_{rz} - N_{c_{rz}},
	\end{equation}
	\begin{equation}\tag{\ref{eq:asymptotic}$f$}
	\sigma \partial_{\bar \tau} c_{zz} = -16 r u_{r} - 4\sigma r u_{z}' + 16 r^2 \partial_{\bar z} u_{z} - 4 r c_{rz} + r^2 \partial_{\bar z} c_{zz}-\sigma c_{zz} - N_{c_{zz}},
	\end{equation}
\end{subequations}
where the nonlinear terms are
\begin{subequations}\label{eq:asymptotic_nonlin}
\begin{equation}
N_{u_{z}} = u_{r} u'_{z} + u_{z} {\partial_{\bar z}} u_{z}, N_{c_{rz}} = u_{r} c'_{rz} + u_{z} {\partial_{\bar z}} c_{rz} - \sigma c_{rr} u'_{z} - c_{rz} {\partial_{\bar z}} u_{z} - c_{rz} u'_{r} - c_{zz} {\partial_{\bar z}} u_{r},
\end{equation}
\begin{equation}
N_{c_{rr}}=u_{r} c'_{rr} + u_{z}{\partial_{\bar z}} c_{rr} - 2 c_{rr} u'_{r} - 2\sigma^{-1}c_{rz} {\partial_{\bar z}} u_{r},N_{c_{zz}}=u_{r} c'_{zz} + u_{z} {\partial_{\bar z}} c_{zz} - 2 c_{rz} u'_{z} - 2 c_{zz} {\partial_{\bar z}} u_{z},
\end{equation}
\end{subequations}
with no-slip boundary conditions applied at the pipe wall $r=1$ and symmetric conditions enforced at the pipe axis $r=0$, i.e., $u_r(1)=u_z(1)=0$ and $u_r(0)=u_z'(0)=0$.
In \eqref{eq:asymptotic} and \eqref{eq:asymptotic_nonlin}, the $O(\sigma)$ and $O(\sigma^2)$ terms are retained. If they are neglected in the bulk region, then we would obtain a solution which is singular at both $r=1$ and $r=0$. Thus, a wall layer and a central layer must be taken into account, being similar to the asymptotic structure in \cite{Dong2022Asymptotic}. Such an unstable mode is found to be possible only when $E\sim \sigma^{-1}$, corresponding to the long-wavelength centre mode for $Re\gg Re_c$ in \cite{Dong2022Asymptotic}. However, when $Re$ is close to  $Re_c$, the three layers merge together, and so the $O(\sigma)$ and $O(\sigma^2)$ terms are kept. It should be noted that now we only have $\beta$, $E$ and $\sigma$ as control parameters in \eqref{eq:asymptotic}, compared to $\beta, Wi, Re, \alpha$ in the original system \eqref{eq:original_compact_linear_eigenvalue}.
Introducing $\boldsymbol{\gamma}=(u_{r},u_{z},p,c_{rr},c_{rz},c_{zz})^T$, the compact form of the asymptotic system \eqref{eq:asymptotic} now reads
\begin{equation}\label{eq:compact-form-taubar}
\left( \boldsymbol{M} {\partial_{\bar \tau}} - \boldsymbol{L} \right) \boldsymbol{\gamma} = \boldsymbol{N}.
\end{equation}
Here, $\boldsymbol{M}$, $\boldsymbol{L}$ and $\boldsymbol{N}$ can be easily deduced by matching equation \eqref{eq:compact-form-taubar} with equation \eqref{eq:asymptotic} and thus are not shown here.

\subsection{Weakly nonlinear analysis of the asymptotic equation system}\label{sec:weakly_nonlinear}

Following \cite{Wan2021Subcritical}, we perform a standard multiple-scale expansion of the reduced equation system \eqref{eq:compact-form-taubar}. To this end, the following expansions are applied as a series of a small quantity $\delta$
\begin{subequations}\label{eq:expansions}
\begin{align}
{\partial_{\bar \tau}} &={\partial_{\bar\tau_{0}}} + \delta {\partial_{\bar\tau_{1}}} + \delta^{2} {\partial_{\bar\tau_{2}}} + O(\delta^{3}), \ \ \ \ \ \ {\partial_{\bar z}} = {\partial_{\bar z_{0}}} + \delta {\partial_{\bar z_{1}}} + O(\delta^{2}),\tag{\ref{eq:expansions}$a,b$}\\
\bgamma &= \delta \bgamma_{1} + \delta^2 \bgamma_{2} + \delta^3 \bgamma_{3} +  O(\delta^{4}), \ \ \ \ \ \   E=E_c - E_c \delta^2 + O(\delta^4),\tag{\ref{eq:expansions}$c,d$}
\end{align}
\end{subequations}
where $E_c= Wi/Re_c$ is the linear critical elasticity number. We use the notation $\delta$ to differentiate the present expansion from that in \cite{Wan2021Subcritical} where the small expansion parameter is $\epsilon$ for the original equation system. The corresponding variables in that paper will be added with a subscript "origi" in the following discussion. Their Eq. (2.10) is copied below
\begin{subequations}\label{eq:expansions_origi}  
	\begin{align}
	{\partial_{t}} &={\partial_{t_{0,\text{origi}}}} + \epsilon {\partial_{t_{1,\text{origi}}}} + \epsilon^{2} {\partial_{t_{2,\text{origi}}}} + O(\epsilon^{3}), \ \ \ {\partial_{z}} = {\partial_{z_{0,\text{origi}}}} + \epsilon {\partial_{z_{1,\text{origi}}}} + O(\epsilon^{2}),\tag{\ref{eq:expansions_origi}$a,b$}\\
	\bgamma &= \epsilon \bgamma_{1,\text{origi}} + \epsilon^2 \bgamma_{2,\text{origi}} + \epsilon^3 \bgamma_{3,\text{origi}} +  O(\epsilon^{4}), \ \ \ Re=Re_c + \epsilon^2 + O(\epsilon^4).\tag{\ref{eq:expansions_origi}$c,d$}
	\end{align}
\end{subequations}
The relation between these two  expansion methods is briefly explained as follows.
From (\ref{eq:expansions_origi}$d$), we can get equivalently $\frac{1}{Re}=\frac{1}{Re_c} - \frac{1}{Re_c^2}\epsilon^{2} + O(\epsilon^{4})$. Multiplying this expansion with $Wi$ leads to
\begin{equation}\label{eq:epsilon_expansion}
\frac{Wi}{Re}=\frac{Wi}{Re_c} - \frac{Wi}{Re_c^2}\epsilon^{2} + O(\epsilon^{4}), \, \, \, \, \text{i.e.,} \, \, \, \, E=E_c - E_c\frac{1}{Re_c}\epsilon^{2} + O(\epsilon^{4}).
\end{equation}
Comparing the expansion \eqref{eq:epsilon_expansion} with the present expansion (\ref{eq:expansions}$d$) shows that these two parameter expansion methods are related by
\begin{equation}\label{eq:delta_epsilon}
\delta^2 = Re_c^{-1} \epsilon^{2}.
\end{equation}

The operators in \eqref{eq:compact-form-taubar} which depend on $\bar\tau$ and $\bar z$ are expanded as $
\boldsymbol{L} = \boldsymbol{L}_{0} + \delta \boldsymbol{L}_{1} + \delta^{2} \boldsymbol{L}_{2} + O(\delta^{3}), \boldsymbol{N} = \delta^{2} \boldsymbol{N}_{2} + \delta^{3} \boldsymbol{N}_{3} + O(\delta^{4})$. Plugging these expansions along with those in equation \eqref{eq:expansions} into \eqref{eq:compact-form-taubar}, and collecting terms on the same order, a series of equations can be obtained,
\begin{subequations}\label{eq:epsilons}
\begin{equation}\label{eq:epsilon1}
\left( \boldsymbol{M} {\partial_{\bar \tau_{0}}} - \boldsymbol{L}_{0} \right) \bgamma_{1} = \boldsymbol{0} \ \text{at order} \ \delta,
\end{equation}
\begin{equation}\label{eq:epsilon2}
\left( \boldsymbol{M} {\partial_{\bar \tau_{0}}} - \boldsymbol{L}_{0} \right) \bgamma_{2} = \left( \boldsymbol{L}_{1} - \boldsymbol{M} {\partial_{\bar \tau_{1}}} \right) \bgamma_{1} + \boldsymbol{N}_{2} \ \text{at order} \ \delta^2,
\end{equation}
\begin{equation}\label{eq:epsilon3}
\left( \boldsymbol{M} {\partial_{\bar \tau_{0}}} - \boldsymbol{L}_{0} \right) \bgamma_{3} = \left( \boldsymbol{L}_{2} - \boldsymbol{M} {\partial_{\bar \tau_{2}}} \right) \bgamma_{1} + \left( \boldsymbol{L}_{1} - \boldsymbol{M} {\partial_{\bar \tau_{1}}} \right) \bgamma_{2} + \boldsymbol{N}_{3} \ \text{at order} \ \delta^3.
\end{equation}
\end{subequations}
Their solutions are assumed to take the wake-like form
\begin{subequations}\label{eq:gammas}
\begin{equation}\label{eq:gamma1_2}
\bgamma_{1} = A_1\tilde{\bgamma}_1 E_{\text{w}} + c.c., \ \ \ \bgamma_{2} = |A_1|^2\tilde{\bgamma}_{20} + ({ \partial_{\bar{z}_1}}A_1 \tilde{\bgamma}_{21}E_{\text{w}} + c.c.) + (A_1^2 \tilde{\bgamma}_{22} E_{\text{w}}^2 + c.c.),
\end{equation}
\begin{equation}
\bgamma_{3} = (A_1\tilde{\bgamma}_{31,1} +{\partial_{\bar{z}_1 \bar{z}_1}}A_1 \tilde{\bgamma}_{31,2} + |A_1|^2 A_1\tilde{\bgamma}_{31,3})E_{\text{w}} + c.c. + \cdots,
\end{equation}
\end{subequations}
where $E_{\text{w}}=\text{exp}({\ri \bar{z}_{0} - \ri c_1 \bar{\tau}_{0}})$, $\tilde{\bgamma}$ with subscripts are the eigenfunctions, and $A_1=A_1(\bar{z}_{1},\bar{\tau}_{1},\bar{\tau}_{2})$ is the complex amplitude of the leading-order wave $\bgamma_{1}$.  
We use $A$ to denote the amplitude of the total disturbance $\bgamma$. Then from the expansion (\ref{eq:expansions}$c$) we know that (to the leading order)
\begin{equation}\label{eq:A_A1_asymp}
A\approx \delta A_1.
\end{equation}
Similarly, from (\ref{eq:expansions_origi}$c$) one obtains $A\approx \epsilon A_{1,\text{origi}}$.

Using \eqref{eq:gammas} in \eqref{eq:epsilons} leads to a set of equations (and their complex conjugates which are omitted here) to be solved in the spectral space
\begin{subequations}\label{eq:spec-epsilons}
	\begin{equation}\label{eq:spec-epsilon1}
	\left( {-\ri c_1} \widetilde{\boldsymbol{M}} - \widetilde{\boldsymbol{L}}_{0}^{(1)} \right) \tilde{\bgamma}_1 = {\bf 0},
	\end{equation}
	\begin{equation}\label{eq:spec-epsilon21}
	{ \partial_{\bar{z}_1}}A_1 \left({-\ri c_1} \widetilde{\boldsymbol{M}} - \widetilde{\boldsymbol{L}}_{0}^{(1)} \right) \tilde{\bgamma}_{21}= \left( \widetilde{\boldsymbol{L}}_{1}^{\circ} { \partial_{\bar{z}_1}}  - \widetilde{\boldsymbol{M}} { \partial_{\bar{\tau}_1}} \right) A_1 \tilde{\bgamma}_1,
	\end{equation}
	\begin{equation}\label{eq:spec-epsilon20-22}
	\left[{(-\ri c_1 + \ri c_1^*)} \widetilde{\boldsymbol{M}} - \widetilde{\boldsymbol{L}}_{0}^{(0)} \right] \tilde{\bgamma}_{20} =  \widetilde{\boldsymbol{N}}_{20},  \left({-2\ri c_1} \widetilde{\boldsymbol{M}} - \widetilde{\boldsymbol{L}}_{0}^{(2)} \right) \tilde{\bgamma}_{22}  =  \widetilde{\boldsymbol{N}}_{22},
	\end{equation}	
	\begin{align}\label{eq:spec-epsilon3}
	&\left({-\ri c_1} \widetilde{\boldsymbol{M}} - \widetilde{\boldsymbol{L}}_{0}^{(1)} \right)\left(A_1\tilde{\bgamma}_{31,1} +{\partial_{\bar{z}_1 \bar{z}_1}}A_1 \tilde{\bgamma}_{31,2} + |A_1|^2 A_1\tilde{\bgamma}_{31,3}\right) =  \notag \\ &\left( \widetilde{\boldsymbol{L}}_{2E} - \widetilde{\boldsymbol{M}} { \partial_{\bar{\tau}_2}} \right) A_1 \tilde{\bgamma}_1  
	+ \left( \widetilde{\boldsymbol{L}}_{1}^{\circ} { \partial_{\bar{z}_1}} - \widetilde{\boldsymbol{M}} { \partial_{\bar{\tau}_1}} \right) { \partial_{\bar{z}_1}}A_1 \tilde{\bgamma}_{21} + |A_1|^2 A_1 \widetilde{\boldsymbol{N}}_{31},
	\end{align}
\end{subequations}
where the superscript ${}^*$ denotes the complex conjugate, and the explicit expressions of the various operators are given in the appendix \ref{Appendix:operators}.

Equation \eqref{eq:spec-epsilon1} with homogeneous boundary conditions forms an eigenvalue problem.  Equations \eqref{eq:spec-epsilon20-22} with nonsingular linear operators are readily solvable. In order to ensure solutions for \eqref{eq:spec-epsilon21} and \eqref{eq:spec-epsilon3}, the solvability conditions must be enforced to eliminate the secular terms on their right-hand sides. To do this, the adjoint of the linear problem \eqref{eq:spec-epsilon1} is introduced as (based on the inner product defined in \cite{Wan2021Subcritical} via integration by part \citep{Luchini2014Adjoint})
\begin{equation}\label{eq:adjoint}
\left({\ri c_1^*} \widetilde{\boldsymbol{M}}^{\dagger}  - \widetilde{\boldsymbol{L}}_{0}^{(1){\dagger}}\right) \tilde{\bgamma}_1^{\dagger} = \boldsymbol{0},
\end{equation}
with $\widetilde{\boldsymbol{M}}^{\dagger}$ and $\widetilde{\boldsymbol{L}}_{0}^{(1){\dagger}}$ described in appendix \ref{Appendix:operators}. Then the solvability condition applied to \eqref{eq:spec-epsilon3} leads to
\begin{equation}\label{eq:GL-tau2 scale}
{\partial_{\bar \tau_{2}}}A_1 = a_{1}A_1 + a_{2}{\partial_{\bar{z}_{1} \bar{z}_{1}}}A_1 + a_{3}{|A_1|}^{2}A_1,
\end{equation}
where the coefficients are
\begin{gather}
a_{1} = \frac{{\langle \widetilde{\boldsymbol{L}}_{2E}\tilde{\bgamma}_1, \tilde{\bgamma}_{1}^{\dagger} \rangle}_{s}}{{\langle \widetilde{\boldsymbol{M}}\tilde{\bgamma}_1, \tilde{\bgamma}_{1}^{\dagger} \rangle}_{s}}, a_{2} = \frac{{\langle (\widetilde{\boldsymbol{L}}_{1}^{\circ} + c_{g}\widetilde{\boldsymbol{M}})\tilde{\bgamma}_{21}, \tilde{\bgamma}_{1}^{\dagger} \rangle}_{s}}{{\langle \widetilde{\boldsymbol{M}}\tilde{\bgamma}_1, \tilde{\bgamma}_{1}^{\dagger} \rangle}_{s}},c_{g} = -\frac{ {\langle \widetilde{\boldsymbol{L}}_{1}^{\circ} \tilde{\bgamma}_{1}, \tilde{\bgamma}_{1}^{\dagger} \rangle}_{s}}{{\langle \widetilde{\boldsymbol{M}}\tilde{\bgamma}_1, \tilde{\bgamma}_{1}^{\dagger} \rangle}_{s}},a_{3} = \frac{{\langle \widetilde{\boldsymbol{N}}_{31}, \tilde{\bgamma}_{1}^{\dagger} \rangle}_{s}}{{\langle \widetilde{\boldsymbol{M}}\tilde{\bgamma}_1, \tilde{\bgamma}_{1}^{\dagger} \rangle}_{s}}.
\end{gather}
The operation ${\langle \tilde{\boldsymbol{f}},\tilde{\boldsymbol{g}} \rangle}_{s} = \int_{0}^{1} \tilde{\boldsymbol{f}} \cdot \tilde{\boldsymbol{g}}^* r \, dr$ is an inner product defined in the spectral space. The derived equation \eqref{eq:GL-tau2 scale} is the third-order Ginzburg-Landau equation (GLE) for the asymptotic reduced equations in the large-$Wi$ limit, to be compared with the GLE for the original equations derived in \cite{Wan2021Subcritical}. Of particular interest is the first Landau coefficient $a_3 = a_{3r} + \ri a_{3i}$ whose real part $a_{3r}$ indicates the primary bifurcation of the laminar flow around its linear critical conditions; its sign being positive (negative) denotes a subcritical (supercritical) bifurcation.
To facilitate the comparison and discussion regarding the GLE, we re-write the GLE for the original equations from Eq. (2.24) of \cite{Wan2021Subcritical} (note that $A$ in \cite{Wan2021Subcritical} is denoted as $A_{1\text{,origi}}$ here)
\begin{equation}\label{eq:GL_t2_scale_origi}
{\partial_{t_{2,\text{origi}}}}A_{1,\text{origi}} = a_{1,\text{origi}}A_{1,\text{origi}} + a_{2,\text{origi}}{\partial_{{z}_{1,\text{origi}} {z}_{1,\text{origi}}}}A_{1,\text{origi}} + a_{3,\text{origi}}{|A_{1,\text{origi}}|}^{2}A_{1,\text{origi}},
\end{equation}
where the time scale ${t_{2,\text{origi}}}$ and space scale ${z}_{1,\text{origi}}$ have been given in (\ref{eq:expansions_origi}$a,b$).
{This equation can also be expressed in time scale $t$, space scale $z$ with amplitude $A$ by inserting the transformation $\partial_{ t}\approx \epsilon^2 \partial_{t_{2,\text{origi}}}$, ${\partial_{zz}} \approx \epsilon^{2} {\partial_{z_{1,\text{origi}}z_{1,\text{origi}}}}$ and that for the disturbance amplitude $A\approx \epsilon A_{1,\text{origi}}$ into \eqref{eq:GL_t2_scale_origi} as
\begin{equation}\label{eq:GL_Atz_origi}
{\partial_{t}}A = \epsilon^2 a_{1,\text{origi}}A + a_{2,\text{origi}}{\partial_{{z}{z}}}A + a_{3,\text{origi}}{|A|}^{2}A.
\end{equation}}

An important issue regarding the evaluation of $a_3$ in \eqref{eq:GL-tau2 scale} is its uniqueness, i.e., $\tilde{\bgamma}_1$ should be normalised to make $a_3$ uniquely determined \citep{Herbert1980}. We follow the normalisation method in \cite{Wan2021Subcritical}, where the linear eigenfunction is normalised so that the square root of the total disturbance energy (kinetic energy plus elastic energy) equals one (see Appendix \ref{Appendix}). 
Therefore, the disturbance amplitude $|A|$ (or equivalently $\delta |A_1|$ as in \eqref{eq:A_A1_asymp}) in our expansion has the physical meaning of square root of the total disturbance energy. 
We hope that this consideration may facilitate comparisons of our results to experiments in the future.

At the end of this section, we would like to discuss the limit of the Oldroyd-B model adopted in this work. The Oldroyd-B model  allows for an infinite extension of polymers and does not account for the shear-thinning effects which can be significant at high $Wi$. The more realistic FENE-P model (finitely extensible nonlinear elastic model with Peterlin closure) overcomes these drawbacks. In the FENE-P model, a new parameter $L_{\text{max}}$ characterising the maximum statistical finite extensibility of polymers is introduced; the model reduces to the Oldroyd-B model when $L_{\text{max}} \to \infty$. In viscoelastic pipe flows, the centre mode instability may disappear when $L_{\text{max}}$ is sufficiently small as illustrated in table 1 of \cite{Zhang2021}. Without a linear critical condition, the multiple-scale expansion method ceases to work. An alternative is the amplitude expansion method enabling a weakly nonlinear expansion of the disturbance around a weakly damped mode (instead of a neutral mode for the multiple-scale expansion) as in \cite{Meulenbroek2003Intrinsic} (such a scenario is also coined bifurcation from infinity). It would be interesting to see whether scaling laws to be presented exist in the results of the amplitude expansion method.

\section{Numerical method and Results}\label{numericalmethod_results}

\subsection{Numerical method and code validation}

The reduced equations including \eqref{eq:spec-epsilon1} and its adjoint, \eqref{eq:spec-epsilon21} and \eqref{eq:spec-epsilon20-22} are solved using a spectral collocation method. We avoid placing a grid point at $r=0$ following \cite{Mohseni2000} and construct the differentiation matrices using the even-odd properties of the variables \citep{Trefethen2000Spectral}, i.e., $u_r$ and $c_{rz}$ are odd and $u_z$, $p$, $c_{rr}$ and $c_{zz}$ even for the axisymmetric mode. Validation of solving the original equations (in linear and weakly nonlinear phases) can be found in \cite{Wan2021Subcritical}. {Computations of the reduced asymptotic equations have been verified by comparing the results with those from the original equations as supplemented in Appendix \ref{Appendix:validation}.}

\subsection{Determining the linear critical conditions}

\begin{figure}
	\centering
	\includegraphics[width=0.29\textwidth,trim= 30 0 70 0,clip]{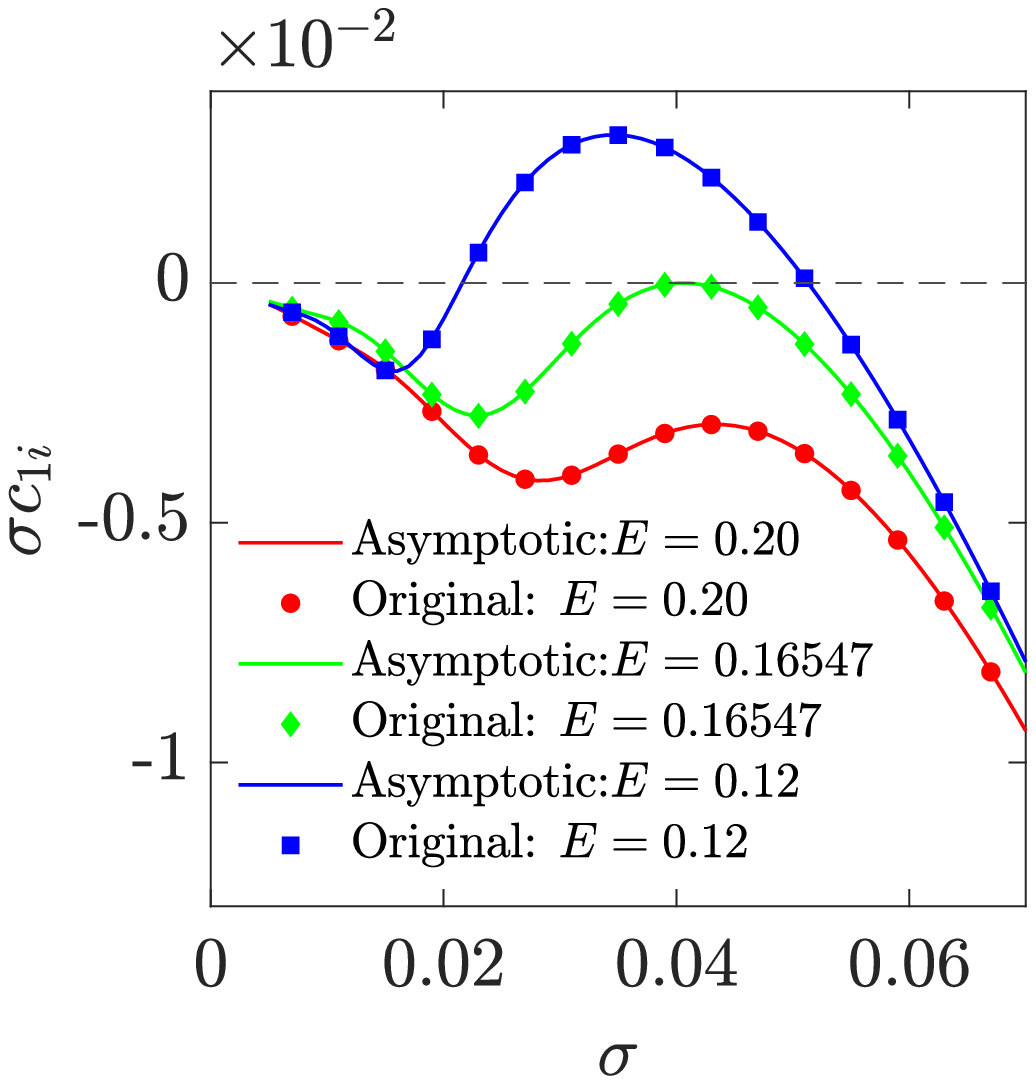}
	\put(-110,97){$(a)$}
	\includegraphics[width=0.35\textwidth,trim= 0 0 40 0,clip]{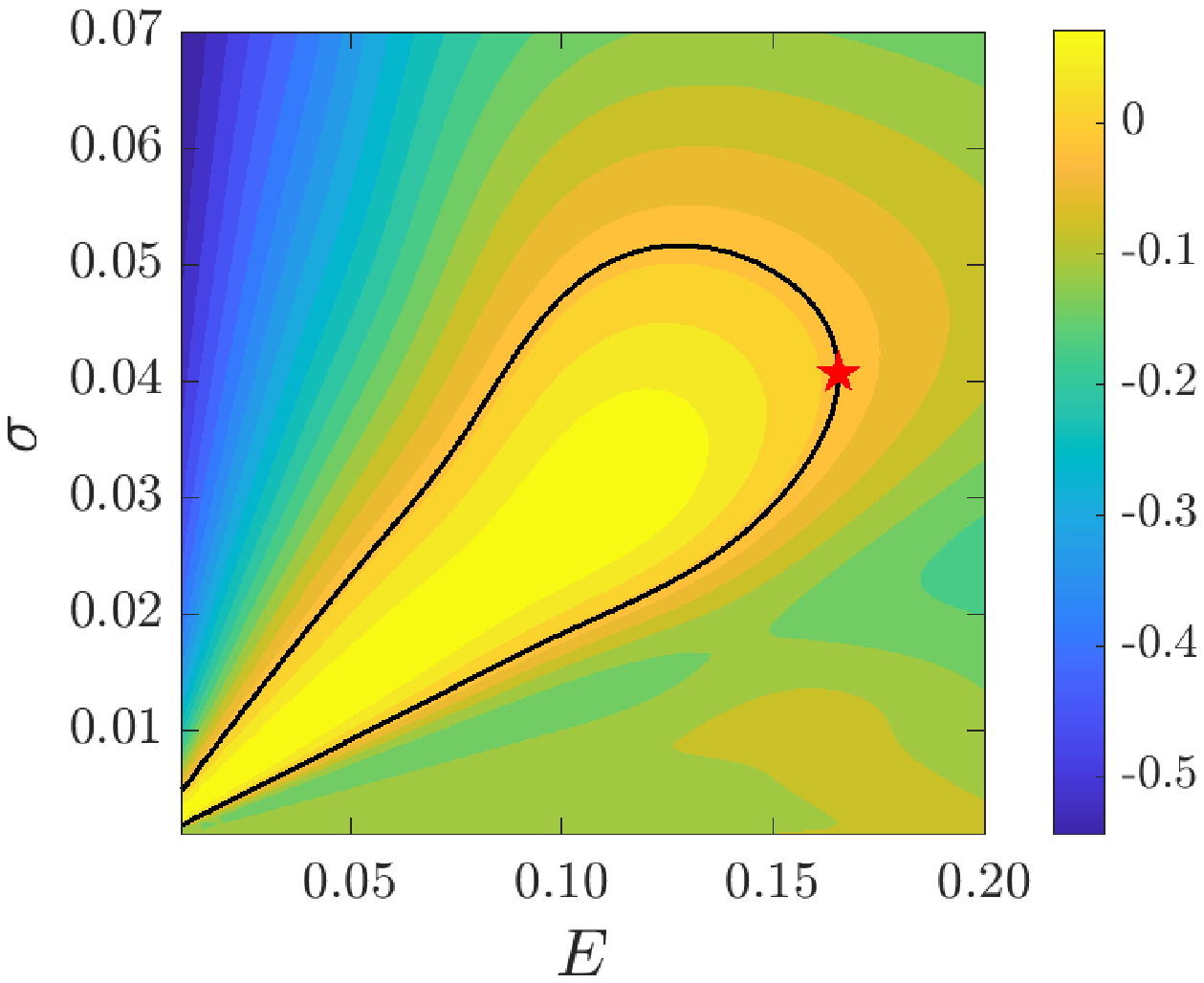}
	\put(-135,97){$(b)$}
	\includegraphics[width=0.35\textwidth,trim= 20 0 10 0,clip]{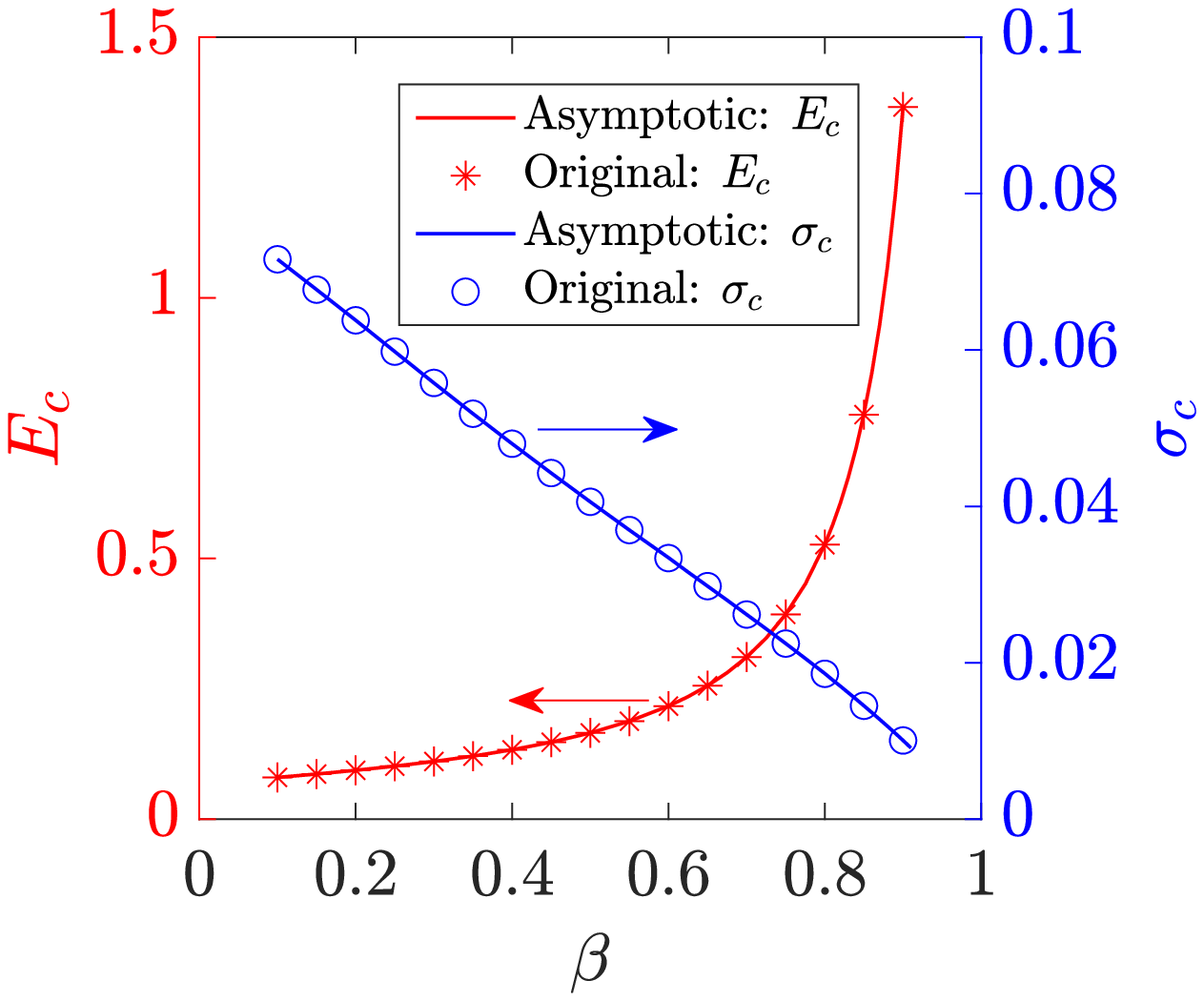}
	\put(-130,97){$(c)$}
	\caption{{$(a)$ Comparison of $\sigma c_{1i}$ as a function of $\sigma$ at different $E$ and $\beta=0.5$. Results from the original equations are obtained at a large $Wi=650$ and then converted according to $\sigma c_1=\omega/\alpha -1$. } $(b)$ Contours of the imaginary part of $c_1$ (i.e., $c_{1i}$) in a $(\sigma,E)$ plane at $\beta=0.5$ with the black line being the neutral curve and the red star marking the linear critical condition. $(c)$ Linear critical conditions $E_c$ and $\sigma_c$ as functions of $\beta$. {Note that in panel $(c)$ $Re_c$ and $\alpha_c$ are first obtained from the original equations at a large $Wi=650$ and then converted to $E_c=Wi/Re_c$ and $\sigma_c=1/(\alpha_c Wi)$ for the comparison with the asymptotic results.}}
	\label{Fig:neutral_curve}
\end{figure}

The multiple-scale expansion is commonly performed around linear critical conditions, which guarantees the convergence of the expansion \citep{Fujimura1989Equivalence}. Next we show how to determine these conditions.
{Figure \ref{Fig:neutral_curve}$(a)$ presents the linear growth rate of the most dangerous mode in the linear asymptotic system (\ref{eq:spec-epsilon1}) for a wide range of $\sigma$ at $\beta=0.5$. As we can see, a good agreement of the results from the two equation systems is achieved. Moreover, each curve shows a local peak and the peak corresponds to the linear critical condition when $E=0.16547$ ({this critical condition is marked by the red star in panel $(b)$})}.
Figure \ref{Fig:neutral_curve}$(b)$ shows the contours of $c_{1i}$ in the $\sigma$--$E$ plane for $\beta=0.5$, with the black curve being the neutral curve. The red star at the right end of the loop marks the linear critical condition (with $E_c=0.16547$ and $\sigma_c=0.04060$), which agrees well with $Wi/Re_c=0.16543$ and $1/(\alpha_c Wi)=0.04058$ for the original equation system at $\beta=0.5$, $Wi=650$, $Re_c=3929.166947$ and $\alpha_c=0.037909$. Because $E=Wi/Re$ and that in the original equations increasing $Re$ brings out instability, in the asymptotic equations decreasing $E$ renders the flow more unstable from the red star.

One of the advantages of the reduced system is that the number of governing parameters is reduced. Reflected in determining the critical conditions, we see that at a given $\beta$, in the original equations the critical condition needs to be determined at each $Wi$, whereas there is only one critical condition in the $\sigma$--$E$ plane in the asymptotic equations. The variations of $E_c$ and $\sigma_c$ with $\beta$ are plotted in figure \ref{Fig:neutral_curve}$(c)$, where results obtained from the original equations at $Wi=650$ are converted accordingly and then superposed to illustrate the good agreement. We can also see that with increasing $\beta$, the critical $E_c$ increases whereas the critical $\sigma_c$ decreases.

\subsection{Bifurcations and the scaling laws at large $Wi$}\label{sec3p3}
\subsubsection{Explaining the scaling law of the first Landau coefficient}
The GLE \eqref{eq:GL-tau2 scale} governs the evolution of disturbance amplitudes around the linear critical conditions. Based on the original equations, \cite{Wan2021Subcritical} found that both subcritical and supercritical bifurcations exist in axisymmetric viscoelastic pipe flows of Oldroyd-B fluids in a large parameter space, and it is mainly the viscosity ratio $\beta$ (related to polymer concentration) that determines the bifurcation type. They reported a large-$Wi$ scaling law for $a_{3}$ (see their figure 13), which cannot be explained by a simple scaling analysis (as adopted by \cite{Chaudhary2021Linear} to explain some linear scalings in their study). In the following, we use the derived reduced system to illustrate the scaling law.

\begin{figure}
	\centering
	\includegraphics[width=0.33\textwidth,trim= 40 0 70 0,clip]{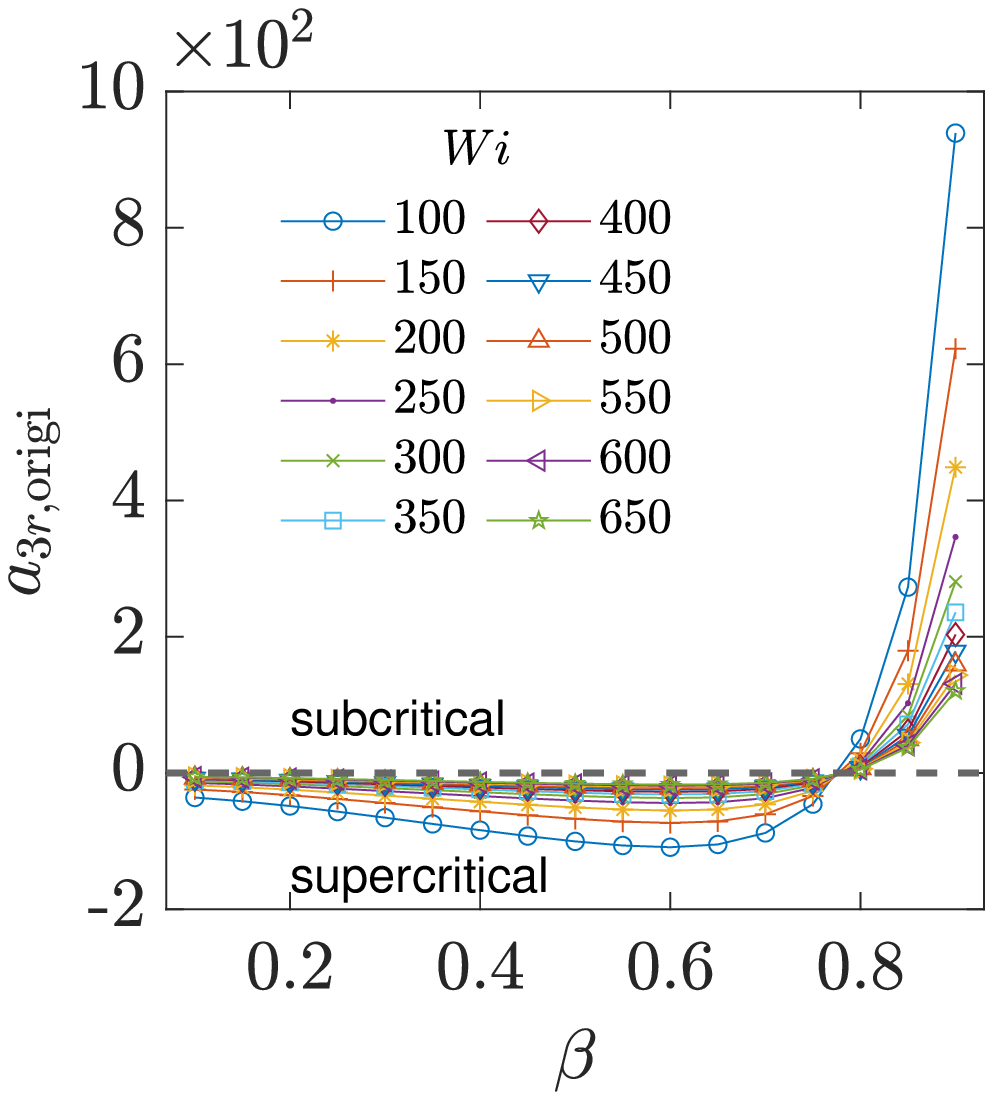}
	\put(-126,112){$(a)$}	\includegraphics[width=0.33\textwidth,trim= 30 0 70 0,clip]{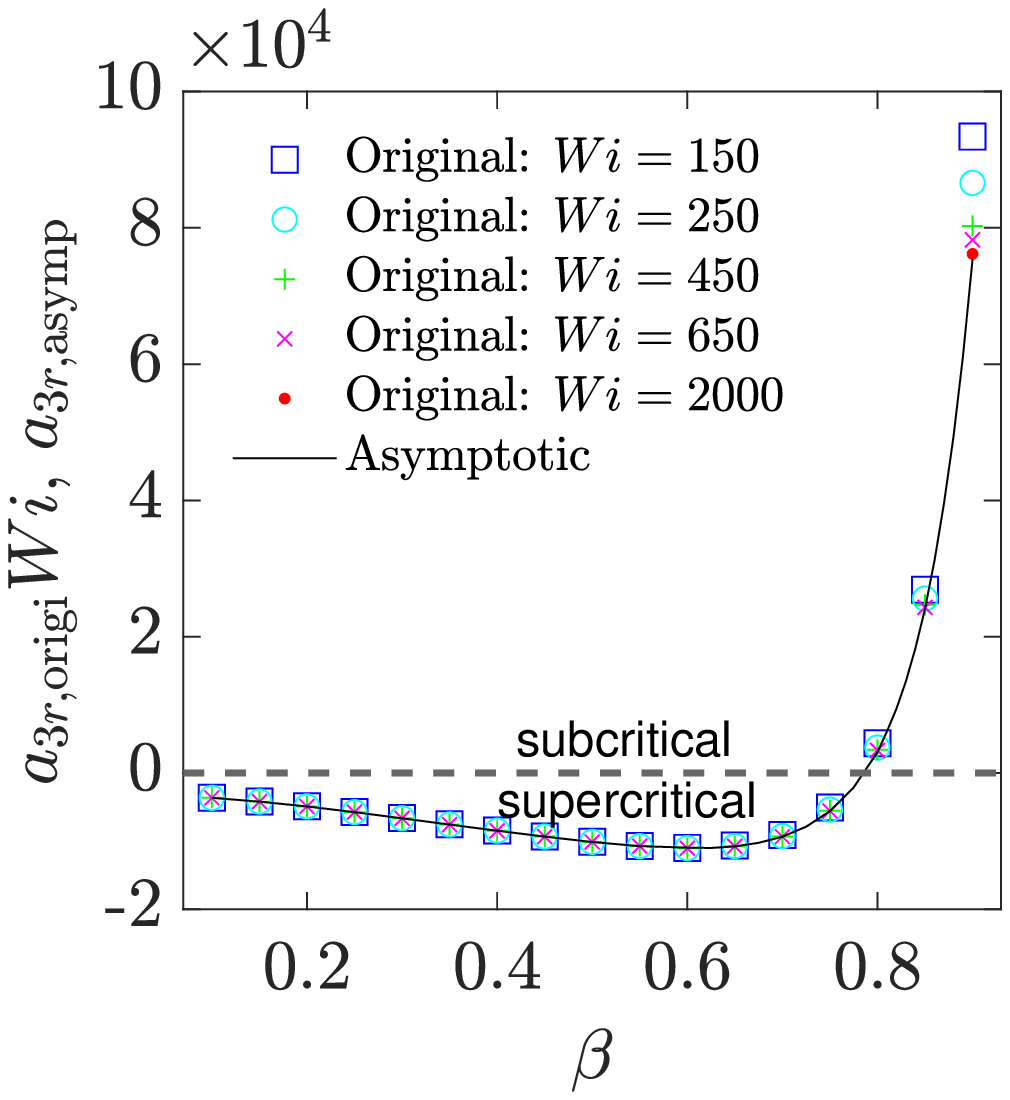}
	\put(-122,112){$(b)$}
	\includegraphics[width=0.33\textwidth,trim= 30 0 70 0,clip]{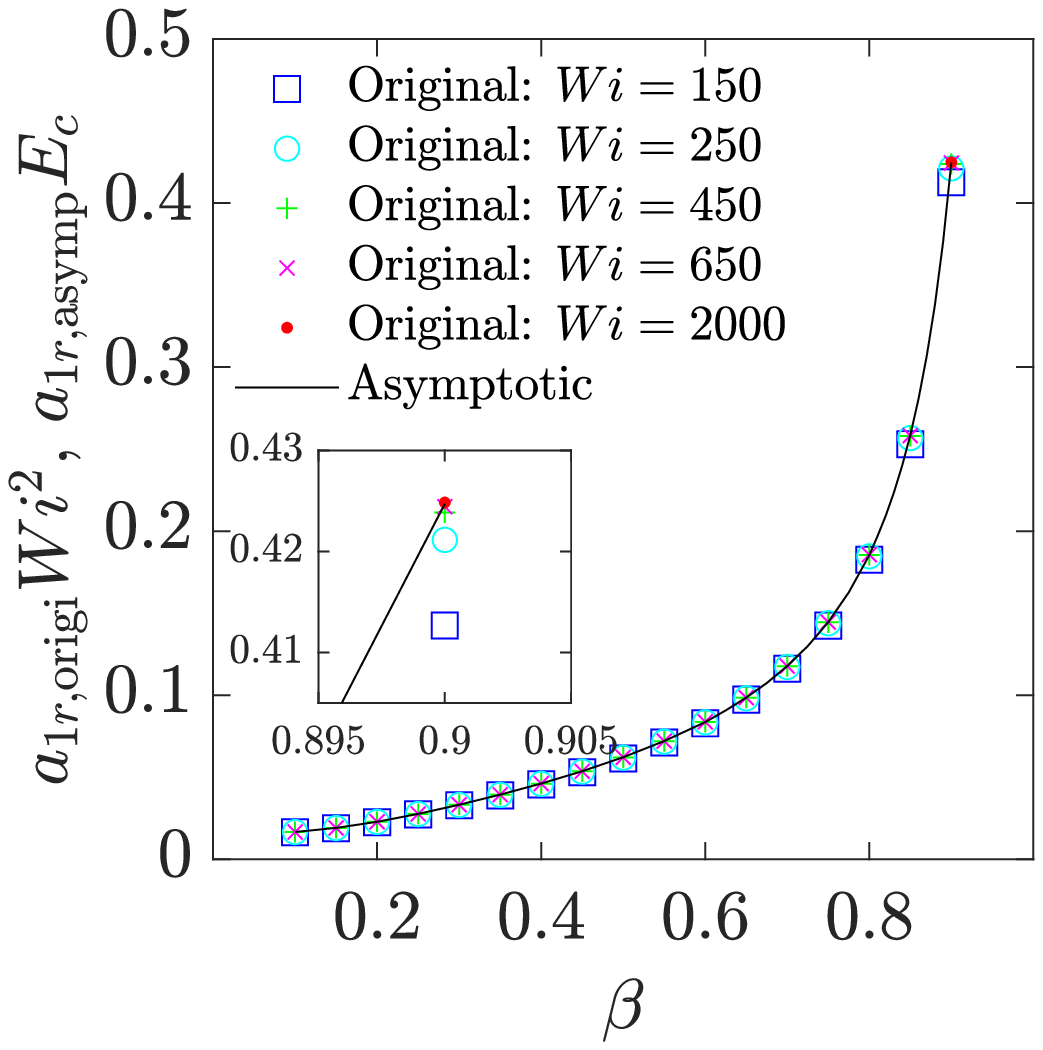}
	\put(-125,112){$(c)$}	
	\caption{Scalings for GLE coefficients in the original equations and in the asymptotic equations. $(a)$ raw data of $a_{3r,\text{origi}}$; $(b)$ scaled data $a_{3r,\text{origi}} Wi$ and results of $a_{3r,\text{asymp}}$ (larger $\beta$ needs higher $Wi$ to accurately present the scaling); $(c)$ scaled data $a_{1r,\text{origi}}Wi^2$ and results of $a_{1r,\text{asymp}}E_c$. The imaginary parts of these coefficients also follow these scalings (not shown). {In panels $(a)$ and $(b)$ data points above the horizontal dashed line correspond to subcritical bifurcations ($a_{3r}>0$) and those below indicate supercritical bifurcations ($a_{3r}<0$).}}
	\label{Fig:a3_a1}
\end{figure}

Figure \ref{Fig:a3_a1}$(a)$ shows the raw data of $a_{3r}$ of the original equations (symbols). After being multiplied by $Wi$ (symbols in panel $(b)$), $a_{3r}$ either exactly collapse on or gradually approach the $a_{3r}$ of the asymptotic equations (black curve in panel $(b)$) when $Wi$ increases, for all the $\beta$ investigated. The advantage of the reduced system is more manifest when $\beta$ is larger, which requires an even larger $Wi$ to present the scaling (see the red dot at $\beta=0.9$).
In the large-$Wi$ limit, the flow bifurcation type is subcritical at large viscosity ratios $\beta$ (small polymer concentration) and changes to be supercritical when $\beta$ is small (large polymer concentration). The bifurcation boundary is $\beta_{\text{crit}}\approx0.785$, which is approximately the same as that in \cite{Wan2021Subcritical} for a large $Wi=650$.
The link between $a_3$ obtained from the asymptotic equations and that from the original equations can be more clearly seen in the multiple-scale expansion of these two sets of equations, as follows. 

Noting that the GLE in \eqref{eq:GL-tau2 scale} is in time scale $\bar{\tau}_2$ and space scale $\bar{z}_1$, our first step to build the link is to convert these scales to time scale $t_2$ and space scale $z_1$. To this end, we introduce the following expansions
\begin{subequations}\label{eq:expansions_t_tau_z}
	\begin{align}
	{\partial_{\tau}} &={\partial_{\tau_{0}}} + \delta {\partial_{\tau_{1}}} + \delta^{2} {\partial_{\tau_{2}}} + O(\delta^{3}), \ \ \ \ \ \ {\partial_{z}} = {\partial_{z_{0}}} + \delta {\partial_{z_{1}}} + O(\delta^{2}), \tag{\ref{eq:expansions_t_tau_z}$a,b$}\\
	{\partial_{\bar t}} &={\partial_{\bar t_{0}}} + \delta {\partial_{\bar t_{1}}} + \delta^{2} {\partial_{\bar t_{2}}} + O(\delta^{3}), \ \ \ \ \ \ \ {\partial_{t}} ={\partial_{t_{0}}} + \delta {\partial_{t_{1}}} + \delta^{2} {\partial_{t_{2}}} + O(\delta^{3}). \tag{\ref{eq:expansions_t_tau_z}$c,d$}
	\end{align}
\end{subequations}
Then, from $\bar t=\alpha t$ in equation \eqref{eq:t_z_tau_bar}, we obtain $\alpha \partial_{\bar t}={\partial_t}$. By comparing the terms of $O(\delta^{2})$ in the expansions of $\partial_{\bar t}$ in $(\ref{eq:expansions_t_tau_z}c)$ and ${\partial_t}$ in $(\ref{eq:expansions_t_tau_z}d)$, we have $\alpha{\partial_{\bar t_{2}}}={\partial_{ t_{2}}}$. From the relations in \eqref{eq:t_z_tau} and \eqref{eq:t_z_tau_bar}, we know that ${\partial_{\bar t}}=-{\partial_{\bar z}} + \sigma {\partial_{\bar \tau}}$; considering the expansions of these partial derivatives in $(\ref{eq:expansions_t_tau_z}c)$ and $(\ref{eq:expansions}a,b)$, we obtain ${\partial_{\bar t_{2}}}=\sigma {\partial_{\bar\tau_{2}}}$ at $O(\delta^{2})$. Therefore, $\partial_{\bar \tau_{2}}=\sigma^{-1}{\partial_{\bar t_{2}}} =\sigma^{-1}\alpha^{-1} \partial_{t_2}= (\alpha Wi)\alpha^{-1} \partial_{t_2} =Wi \partial_{t_2}$. Using this relation in \eqref{eq:GL-tau2 scale} results in (note that for the spatial derivative ${\partial_{\bar{z}_{1} \bar{z}_{1}}}=\alpha^{-2} {\partial_{z_{1} z_{1}}}$ is used)
\begin{equation}\label{eq:GLE_compare}
Wi \partial_{t_2} A_1 = a_{1}A_1 + \alpha^{-2} a_{2} {\partial_{z_{1} z_{1}}}A_1 + a_{3}{|A_1|}^{2}A_1.
\end{equation}

{The second step is to further transform \eqref{eq:GLE_compare} into a GLE in time scale $t$ and space scale $z$ with amplitude $A$. 
Inserting the transformation $\partial_{ t}\approx \delta^2 \partial_{t_{2}}$, ${\partial_{zz}} \approx \delta^{2} {\partial_{z_{1}z_{1}}}$ and that for the disturbance amplitude $A\approx \delta A_1$ into \eqref{eq:GLE_compare} leads to
\begin{equation}\label{eq:GLE_compare_Atz}
\partial_{t} A =Wi^{-1}\delta^{2} a_{1}A + Wi^{-1}\alpha^{-2} a_{2} {\partial_{zz}}A + Wi^{-1}a_{3}{|A|}^{2}A.
\end{equation}
This equation is to be compared with (\ref{eq:GL_Atz_origi}). 
Comparing \eqref{eq:GLE_compare_Atz} with \eqref{eq:GL_Atz_origi} term by term results in (recall the relation in \eqref{eq:delta_epsilon})
\begin{subequations}\label{eq:full-redu-trans}
	\begin{align}
	&a_{1,\text{origi}} =Wi^{-1} Re_c^{-1} a_{1,\text{asymp}} = Wi^{-2} E_c a_{1,\text{asymp}}, \\
	&a_{2,\text{origi}} =Wi^{-1}\alpha^{-2} a_{2,\text{asymp}}, \ \ a_{3,\text{origi}} = Wi^{-1} a_{3,\text{asymp}},\tag{\ref{eq:full-redu-trans}$b,c$}
	\end{align}
\end{subequations}
where the subscript "asymp" is additionally added to denote the coefficients in the asymptotic GLE \eqref{eq:GL-tau2 scale}, \eqref{eq:GLE_compare} and \eqref{eq:GLE_compare_Atz}.}

Since $a_{3,\text{asymp}}$ for a given $\beta$ is fixed at the critical state $(\sigma_c, E_c)$, the relation (\ref{eq:full-redu-trans}$c$) implies $a_{3,\text{origi}}$ is proportional to $Wi^{-1}$, in agreement with the scaling law observed numerically in \cite{Wan2021Subcritical} at large $Wi$.
Thanks to the asymptotic equation system, we are able to easily identify more scaling results in the weakly nonlinear phase (see (\ref{eq:full-redu-trans}$a,b$)). For example, figure \ref{Fig:a3_a1}$(c)$ shows the scaling law for the coefficient $a_1$ in GLE.
We plot $a_{1r,\text{origi}}Wi^2$ and $a_{1r,\text{asymp}} E_c$ for comparison. They are found to agree well when $Wi$ is sufficiently large, confirming the relation (\ref{eq:full-redu-trans}$a$) since $Re_c=E_c^{-1}Wi$ by definition, i.e., $a_{1,\text{origi}}$ is proportional to $Wi^{-2}$.

\subsubsection{Scaling law of the equilibrium amplitude $A_e$}
According to the third-order GLE (Eq. \eqref{eq:GL-tau2 scale} in the present asymptotic analysis and Eq. \eqref{eq:GL_t2_scale_origi} in the original equations), finite-amplitude disturbances are required to trigger the subcritical transition, meaning that the flow is linearly stable but could be nonlinearly unstable. In this case, we can use the equilibrium solution of the GLE to quantify the amplitude threshold beyond which the transition occurs. For supercritical bifurcations, infinitesimal disturbances can cause the flow transition and the third-order nonlinearity stabilises the flow, leading to a saturated state whose amplitude could also be characterised by the equilibrium solution of the GLE. 
Based on the above scaling laws of the coefficients in the GLE, a scaling of the corresponding equilibrium amplitude of disturbance can be derived in the neighbourhood of linear critical conditions either from the original GLE {(not shown here) or from the asymptotic GLE. We illustrate the derivation from the asymptotic GLE as follows.

Starting from the asymptotic $3^{\text{rd}}$-order GLE \eqref{eq:GLE_compare_Atz}, if the diffusion term of the disturbance amplitude is ignored, the equilibrium amplitude $A_e$ (i.e., the modulus of $A$) can be obtained by setting $\partial_t A = 0$ as (note that to calculate $A_e$ only the real parts of the coefficients are used; recall the parameter expansion $E=E_c - E_c \delta^2 + O(\delta^4)$ in equation (\ref{eq:expansions}$d$))
\begin{equation}\label{eq:redu_Ae_Re}
0 = \delta^2 a_{1r} |A| +   a_{3r}{|A|}^{3}  \ \ \ \rightarrow \ \ \  A_e=|A|=\sqrt{-\delta^2 \frac{a_{1r}}{a_{3r}}}=\sqrt{ \frac{a_{1r}}{a_{3r}} \frac{E-E_c}{E_c}}=\sqrt{ \frac{a_{1r}}{a_{3r}} \frac{Re_c-Re}{Re}}.
\end{equation}
Here for supercritical bifurcations $Re>Re_c$, $a_{1r}>0$ and $a_{3r}<0$; for subcritical bifurcations $Re<Re_c$, $a_{1r}>0$ and $a_{3r}>0$. Therefore, $A_e$ is always a real number. This corresponds to the finite-amplitude equilibrium solution (near linear criticality) in the axisymmetric viscoelastic pipe flows calculated using GLE.
Under the assumptions that $Wi \gg 1$ and $E=O(1)$ (thus $Re \gg 1$) and that the above analysis is restricted to the vicinity of the linear critical conditions ($|Re_c - Re|\ll Re_c$), there is the following scaling law of the equilibrium amplitude $A_e$ in terms of $Wi$
\begin{equation}\label{eq:redu_Ae_Wi}
A_e=\sqrt{ \frac{a_{1r}}{a_{3r}} \frac{Re_c-Re}{Re}} \approx \sqrt{ \frac{a_{1r}}{a_{3r}} \frac{Re_c-Re}{Re_c}} = \sqrt{ \frac{ a_{1r} E_c (Re_c-Re)}{a_{3r} Wi} } \propto Wi^{-\frac{1}{2}}.
\end{equation}}

As mentioned in the introduction, in both Newtonian channel and pipe flows, scaling laws of the amplitude threshold have been reported in subcritical regimes and favourable agreements between theoretical predictions \citep{Chapman2002Subcritical,Waleffe2005Transition} and experimental observations \citep{Hof2003Scaling,Philip2007Scaling,Lemoult2012Experimental} have been achieved. Scaling laws in viscoelastic flows have also been reported in literature such as \cite{Jovanovic2010Transient} and \cite{Morozov2007}, both concerning elastic instabilities at vanishing $Re$. The scaling law of the equilibrium amplitude $A_e \propto Wi^{-1/2}$ derived in this work pertains to the EIT transition (in the parameter range that $Wi \gg 1$, $E=O(1)$). This result can be extended to other flow quantities such as the mean-flow distortion ($\delta^2|A_1|^2\tilde{\bgamma}_{20}\approx|A|^2\tilde{\bgamma}_{20}=A_e^2\tilde{\bgamma}_{20} \propto Wi^{-1}$ at equilibrium states), which may be of interest to experimentalists who can measure the disturbance amplitude in either the flow field or the conformation tensor field.

\section{Discussion and Conclusions} \label{Conclusions}

This work derived an asymptotic nonlinear system for the centre mode \citep{Garg2018Viscoelastic} at large $Wi$ in axisymmetric viscoelastic pipe flows. After applying the asymptotic analysis, we reduce the number of parameters from 4 ($\beta$, $Wi$, $Re$ and $\alpha$ in the original system) to 3 ($\beta$, $E$ and $\sigma$ in the reduced system) and the number of unknowns from 7 to 6 (as the component $c_{\theta\theta}$ is decoupled). Detailed comparisons between these two  systems show that the asymptotic equations can well capture the linear and weakly nonlinear characteristics of the flow near linear critical conditions when $Wi$ is large enough.
More importantly, the scaling law $a_3\varpropto Wi^{-1}$ when $Wi$ is large, which is numerically found using the original equations by \cite{Wan2021Subcritical}, can be successfully explained via the multiple-scale expansion of the reduced system, circumventing much numerical difficulty in resolving large-$Wi$ flows and revealing the inherent relations of the Landau coefficients in the two systems. The reduced system also enables us to easily discover and explain more scaling results for $a_1, a_2$ and $A_e$. Particularly, because the equilibrium amplitude $A_e$ of the disturbance around the linear critical conditions scales with $Wi^{-1/2}$, the amplitude of the mean flow distortion follows $A_e^2 \propto Wi^{-1}$.
Future works can consider confirming the scaling laws we identified and searching for nonlinear equilibrium solutions to the reduced asymptotic equations for large-$Wi$ flows.

The current asymptotic analysis exemplifies an approach to studying the large-$Wi$ and large-$Re$ viscoelastic flow (which is often a terrible struggle for conventional methods). In the $Wi$-$Re$ schematic showing various transition routes to EIT (see e.g. \cite{Graham2014Drag,Datta2021,Sanchez2022Understanding}), our work represents a rare probe into the top-right corner of the parameter space, differentiating itself from most of the existing works.
The limitation of this work lies in the assumptions such as the elasticity number $E$ being of order 1, the usage of the simple Oldroyd-B fluid model and the restrictions confined to the neighbourhood of the linear critical conditions.
However, to some degree, it is such assumptions that facilitate the observation of the scaling laws.
In a broader perspective, searching for scaling laws in a fluid system has always been an adventurous and rewarding endeavour for fluid dynamicists. The scaling laws can neatly provide the trend of parametric effects in the experimentally/numerically unavailable regime. Our results extend the linear scaling laws first observed by \cite{Garg2018Viscoelastic} in viscoelastic pipe flows to the nonlinear regime, utilising the theoretical tools developed in \cite{Wan2021Subcritical} and \cite{Dong2022Asymptotic}. We hope that our results will evoke more future work along this direction and contribute to the general understanding of the nonlinear dynamics in viscoelastic flows.

\begin{acknowledgments}
D.W. is supported by a PhD scholarship (No. 201906220200) from the China Scholarship Council and a NUS research scholarship. M.Z. acknowledges the financial support of Ministry of Education, Singapore (with the WBS no. R-265-000-661-112). M.D. is supported by National Science Foundation of China (grant no. U20B2003).
\end{acknowledgments}

Declaration of Interests. The authors report no conflict of interest.

\begin{appendix}
	
\section{Operators in spectral space for the multiple-scale expansion}\label{Appendix:operators}
The various operators appearing in \eqref{eq:spec-epsilons} are described as follows. The weight matrix $\widetilde{\boldsymbol{M}}=\mbox{diag}[0,\sigma,0,\sigma,\sigma,\sigma]$, and $\widetilde{\boldsymbol{L}}_{0}^{(l)}$ with $l=(0,1,2)$ can be deduced from the linearised equations of \eqref{eq:asymptotic} by replacing $\partial_{\bar z}$ there with $l\ri$. The non-zero elements of $\widetilde{\boldsymbol{L}}_{1}^{\circ}$ are 
\begin{gather}
\widetilde{{L}}_{1,22}^{\circ}=r^2,  \
\widetilde{{L}}_{1,23}^{\circ}=-1,  \
\widetilde{{L}}_{1,26}^{\circ}=(1-\beta)E_c,  \ 
\widetilde{{L}}_{1,32}^{\circ}=1,  \ 
\widetilde{{L}}_{1,41}^{\circ}=-4r, \ 
\widetilde{{L}}_{1,44}^{\circ}=r^2, \notag\\
\widetilde{{L}}_{1,51}^{\circ}=8 r^2, \
\widetilde{{L}}_{1,52}^{\circ}=-2\sigma r,  \ 
\widetilde{{L}}_{1,55}^{\circ}=r^2,\
\widetilde{{L}}_{1,62}^{\circ}=16 r^2, \
\widetilde{{L}}_{1,66}^{\circ}=r^2,
\end{gather}
where the two-digit subscript after the comma indexes row and column respectively. The non-zero elements of $\widetilde{\boldsymbol{L}}_{2E}$ are
\begin{equation}
\widetilde{{L}}_{2E,22}=-\sigma \beta E_c (\partial_{rr} + r^{-1} \partial_{r}), \widetilde{{L}}_{2E,25}=-(1-\beta)E_c(\partial_{r}+r^{-1}), \widetilde{{L}}_{2E,26}=-\ri (1-\beta)E_c.
\end{equation}
The nonlinear operators $\widetilde{\boldsymbol{N}}_{20}$, $\widetilde{\boldsymbol{N}}_{22}$ and  $\widetilde{\boldsymbol{N}}_{31}$ in \eqref{eq:spec-epsilons} are given as
\begin{align}\label{eq:nonlinear_N2_N3}
&\widetilde{\boldsymbol{N}}_{20} = \widetilde{\boldsymbol{N}}_f (\tilde{\bgamma}_{1},\tilde{\bgamma}_{1}^*,-1) + \widetilde{\boldsymbol{N}}_f (\tilde{\bgamma}_{1}^*,\tilde{\bgamma}_{1},1), \ \ \ \ \ \ \widetilde{\boldsymbol{N}}_{22} = \widetilde{\boldsymbol{N}}_f (\tilde{\bgamma}_{1},\tilde{\bgamma}_{1},1), \notag\\
&\widetilde{\boldsymbol{N}}_{31} = \widetilde{\boldsymbol{N}}_f (\tilde{\bgamma}_{1},\tilde{\bgamma}_{20},0) + \widetilde{\boldsymbol{N}}_f (\tilde{\bgamma}_{20},\tilde{\bgamma}_{1},1) + \widetilde{\boldsymbol{N}}_f (\tilde{\bgamma}_{1}^*,\tilde{\bgamma}_{22},2) + \widetilde{\boldsymbol{N}}_f (\tilde{\bgamma}_{22},\tilde{\bgamma}_{1}^*,-1).
\end{align}
The function $\widetilde{\boldsymbol{N}}_f$ is defined as $\widetilde{\boldsymbol{N}}_f (\tilde{\boldsymbol{f}}_1,\tilde{\boldsymbol{f}}_2,q)=-(0,\tilde n_{u_z}, 0, \tilde n_{c_{rr}}, \tilde n_{c_{rz}}, \tilde n_{c_{zz}})^T$ with its elements (corresponding to the nonlinear terms in \eqref{eq:asymptotic_nonlin} in spectral space) being
\begin{align}
&\tilde n_{u_z}(\tilde{\boldsymbol{f}}_1,\tilde{\boldsymbol{f}}_2,q) = \tilde{f}_{1,u_r} \tilde{f}'_{2,u_z} + \ri q \tilde{f}_{1,u_z} \tilde{f}_{2,u_z}, \notag\\
&\tilde n_{c_{rr}}(\tilde{\boldsymbol{f}}_1,\tilde{\boldsymbol{f}}_2,q) = \tilde{f}_{1,u_r} \tilde{f}'_{2,c_{rr}} + \ri q \tilde{f}_{1,u_z} \tilde{f}_{2,c_{rr}} - 2 \tilde{f}_{1,c_{rr}} \tilde{f}'_{2,u_r} - 2 \sigma^{-1} \ri q \tilde{f}_{1,c_{rz}} \tilde{f}_{2,u_r}, \notag\\
&\tilde n_{c_{rz}}(\tilde{\boldsymbol{f}}_1,\tilde{\boldsymbol{f}}_2,q) = \tilde{f}_{1,u_r} \tilde{f}'_{2,c_{rz}} + \ri q \tilde{f}_{1,u_z} \tilde{f}_{2,c_{rz}} - \sigma \tilde{f}_{1,c_{rr}} \tilde{f}'_{2,u_z} -  \ri q \tilde{f}_{1,c_{rz}} \tilde{f}_{2,u_z} \notag\\
& \ \ \ \ \ \ \ \ \ \ \ \ \ \ \ \ \ \ \ \ \ - \tilde{f}_{1,c_{rz}} \tilde{f}'_{2,u_r} - \ri q \tilde{f}_{1,c_{zz}} \tilde{f}_{2,u_r}, \notag\\
&\tilde n_{c_{zz}}(\tilde{\boldsymbol{f}}_1,\tilde{\boldsymbol{f}}_2,q) = \tilde{f}_{1,u_r} \tilde{f}'_{2,c_{zz}} + \ri q \tilde{f}_{1,u_z} \tilde{f}_{2,c_{zz}} - 2 \tilde{f}_{1,c_{rz}} \tilde{f}'_{2,u_z} - 2 \ri q \tilde{f}_{1,c_{zz}} \tilde{f}_{2,u_z}.
\end{align}
The second subscripts in $\tilde f_{1,\cdot}$ and $\tilde f_{2,\cdot}$ mark the corresponding components in the column vectors $\tilde{\boldsymbol{f}}_1$ and $\tilde{\boldsymbol{f}}_2$.
	
In the adjoint problem \eqref{eq:adjoint}, $\widetilde{\boldsymbol{M}}^{\dagger}=\widetilde{\boldsymbol{M}}$ (self-adjoint), and the non-zero elements in $\widetilde{\boldsymbol{L}}_{0}^{(1){\dagger}}$ include
\begin{gather}
\widetilde{L}_{0,12}^{(1)\dagger} = 2r, \, \, \, 
\widetilde{L}_{0,13}^{(1)\dagger} = -\partial_r, \,  \, \, 
\widetilde{L}_{0,14}^{(1)\dagger} = -2\sigma\partial_r - 2\sigma r^{-1} + 4\ri r,\notag\\
\widetilde{L}_{0,15}^{(1)\dagger} = 6\sigma - 8 \ri r^2 + 2\sigma r \partial_r, \,  \, \, 
\widetilde{L}_{0,16}^{(1)\dagger} = -16 r, \,  \, \, 
\widetilde{L}_{0,22}^{(1)\dagger} = -\ri r^2 + \sigma \beta E_c (\partial_{rr} + r^{-1}\partial_r),\notag\\
\widetilde{L}_{0,23}^{(1)\dagger} = -\ri, \,  \, \, 
\widetilde{L}_{0,25}^{(1)\dagger} = 2\ri \sigma r - \sigma^2 \partial_r-\sigma^2 r^{-1}, \,  \, \, 
\widetilde{L}_{0,26}^{(1)\dagger} = -16 \ri r^2 + 4 \sigma r \partial_r + 8 \sigma,\notag\\
\widetilde{L}_{0,31}^{(1)\dagger} = \partial_r + r^{-1}, \, \,  \, 
\widetilde{L}_{0,32}^{(1)\dagger} = \ri, \,  \, \, 
\widetilde{L}_{0,44}^{(1)\dagger} = -\ri r^2 - \sigma, \,  \, \, 
\widetilde{L}_{0,45}^{(1)\dagger} = -2\sigma r, \, \, \, 
\widetilde{L}_{0,52}^{(1)\dagger} = -(1-\beta)E_c \partial_r,\notag\\
\widetilde{L}_{0,55}^{(1)\dagger} = -\ri r^2 - \sigma, \,  \, \, 
\widetilde{L}_{0,56}^{(1)\dagger} =-4r, \, \, \, 
\widetilde{L}_{0,62}^{(1)\dagger} = -\ri (1-\beta) E_c, \, \, \, 
\widetilde{L}_{0,66}^{(1)\dagger} = -\ri r^2 - \sigma.
\end{gather}
	
\section{Normalisation of the linear eigenfunction $\tilde{\boldsymbol{\gamma}}_1$}\label{Appendix}
The linear eigenvalue problem in \eqref{eq:spec-epsilon1} can be arbitrarily scaled. To make the third-order Landau coefficient $a_3$ in \eqref{eq:GL-tau2 scale} uniquely determined, $\tilde{\boldsymbol{\gamma}}_1$ should be normalised \citep{Herbert1980}. The normalisation condition used in \cite{Wan2021Subcritical} is (their equation (2.26))
\begin{equation}\label{eq:total_energy_origi}
\sqrt{\frac{1}{2} \int_{0}^{1} \left( \left(|\tilde{u}_{r\text{F}1}|^2 + |\tilde{u}_{z\text{F}1}|^2 \right)  +  \frac{1-\beta}{Re Wi}  \left( |\tilde{g}_{rr\text{F}1}|^{2}+2|\tilde{g}_{rz\text{F}1}|^{2}+|\tilde{g}_{\theta\theta\text{F}1}|^{2}+|\tilde{g}_{zz\text{F}1}|^2\right) \right) \, rdr} = 1.
\end{equation}
We here follow their normalisation condition in that a different normalisation will result in values of $a_3$ that cannot be quantitatively compared with the results there, although the sign of $a_{3r}$ remains unchanged.

It should be noted that in \eqref{eq:total_energy_origi} the parameters $\beta$, $Re$ and $Wi$ are used, while we have parameters $\beta$, $\sigma$ and $E$ in the reduced equation system \eqref{eq:asymptotic}. In order for a comparison, we choose a sufficiently large $Wi=5000$ as our study focuses on $Wi\gg 1$.
With $Wi$ given, at a linear critical condition $(E_c, \sigma_c)$ for a certain $\beta$, the linear critical $Re$ and wavenumber can be obtained as $Re_c=Wi/E_c$ and $\alpha_c=1/(\sigma_c Wi)$. Then the linear eigenfunction $\tilde{\boldsymbol{\gamma}}_1=(\tilde{u}_{r1},\tilde{u}_{z1},\tilde{p}_{1},\tilde{c}_{rr1},\tilde{c}_{rz1},\tilde{c}_{zz1})^T$ in the present analysis can be normalised as
\begin{equation}\label{eq:total_energy_asymp}
\sqrt{\frac{1}{2} \int_{0}^{1} \left( \left(|\alpha_c \tilde{u}_{r1}|^2 + |\tilde{u}_{z1}|^2 \right)  +  \frac{1-\beta}{Re_c Wi}  \left( |\tilde{g}_{rr1}|^{2}+2|\tilde{g}_{rz1}|^{2}+|\tilde{g}_{zz1}|^2\right) \right) \, rdr} = 1.
\end{equation}
Here the usage of the polymer deformation tensor components $\tilde{g}_{rr1}$, $\tilde{g}_{rz1}$ and $\tilde{g}_{zz1}$ follows the geometric decomposition of the conformation tensor $\boldsymbol{c}$ proposed by \cite{Hameduddin2018Geometric} and further developed in \cite{Hameduddin2019Perturbative} and \cite{Hameduddin2019b}. With this geometric decomposition, the elastic energy can be unambiguously defined. These polymer deformation tensor components are calculated according to the relation 

\begin{equation}\label{eq:cg_transformation}
\begin{pmatrix}
\tilde{g}_{rr1} \\ \tilde{g}_{rz1} \\ \tilde{g}_{zz1}
\end{pmatrix} = 
\begin{pmatrix}
C_{rr} & 0 & 0\\
C_{rz} & S & 0 \\
C_{rz}^{2}/C_{rr} & 2SC_{rz}/C_{rr} & S^2/C_{rr}
\end{pmatrix}^{-1} \begin{pmatrix}
\sigma_c^{-1}\tilde{c}_{rr1} \\ \alpha_c^{-1}\sigma_c^{-2}\tilde{c}_{rz1} \\ \alpha_c^{-2}\sigma_c^{-2}\tilde{c}_{zz1}
\end{pmatrix},
\end{equation}
where $S=\sqrt{C_{rr}C_{zz}-C_{rz}^2}$; $C_{rr}=1$, $C_{rz}=Wi U'_z$ and $C_{zz}=1+ 2 Wi^2 U_z'^{2}$. The additional coefficients $\alpha_c$ and $\sigma_c$ in \eqref{eq:total_energy_asymp} and \eqref{eq:cg_transformation} are due to the rescaling process described in \eqref{eq:rescaling}. For a comparison with results from the original equation system, rescaling back is needed. Note that in the above normalisation condition, the conformation tensor component $\tilde{c}_{\theta\theta1}$ (and so $\tilde{g}_{\theta\theta1}$) does not appear any longer.

\section{Numerical validation by comparing with the original system}\label{Appendix:validation}

\begin{figure}
	\centering
	\includegraphics[width=0.33\textwidth,trim= 30 0 67 0,clip]{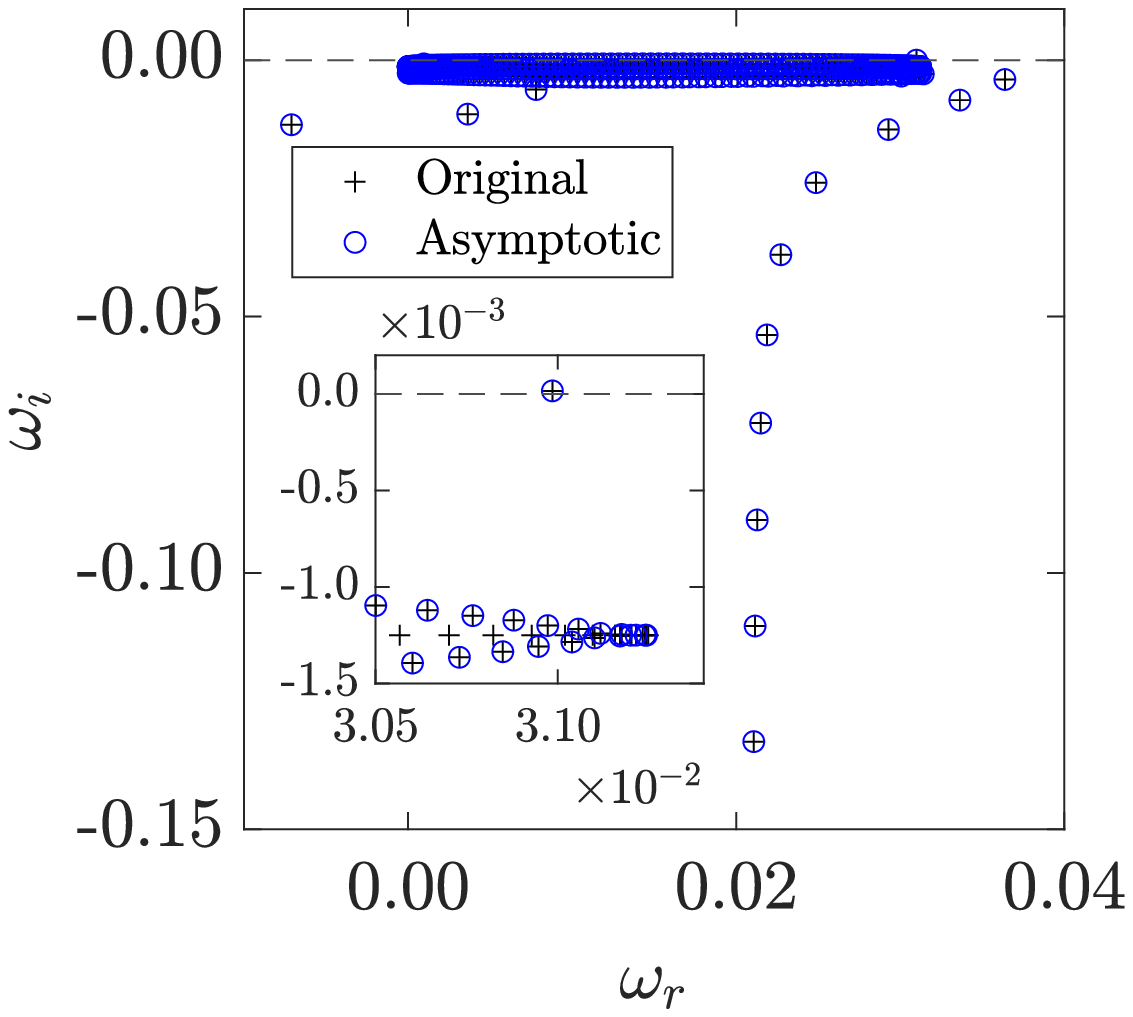}
	\put(-130,107){$(a)$}
	\includegraphics[width=0.33\textwidth,trim= 30 0 70 0,clip]{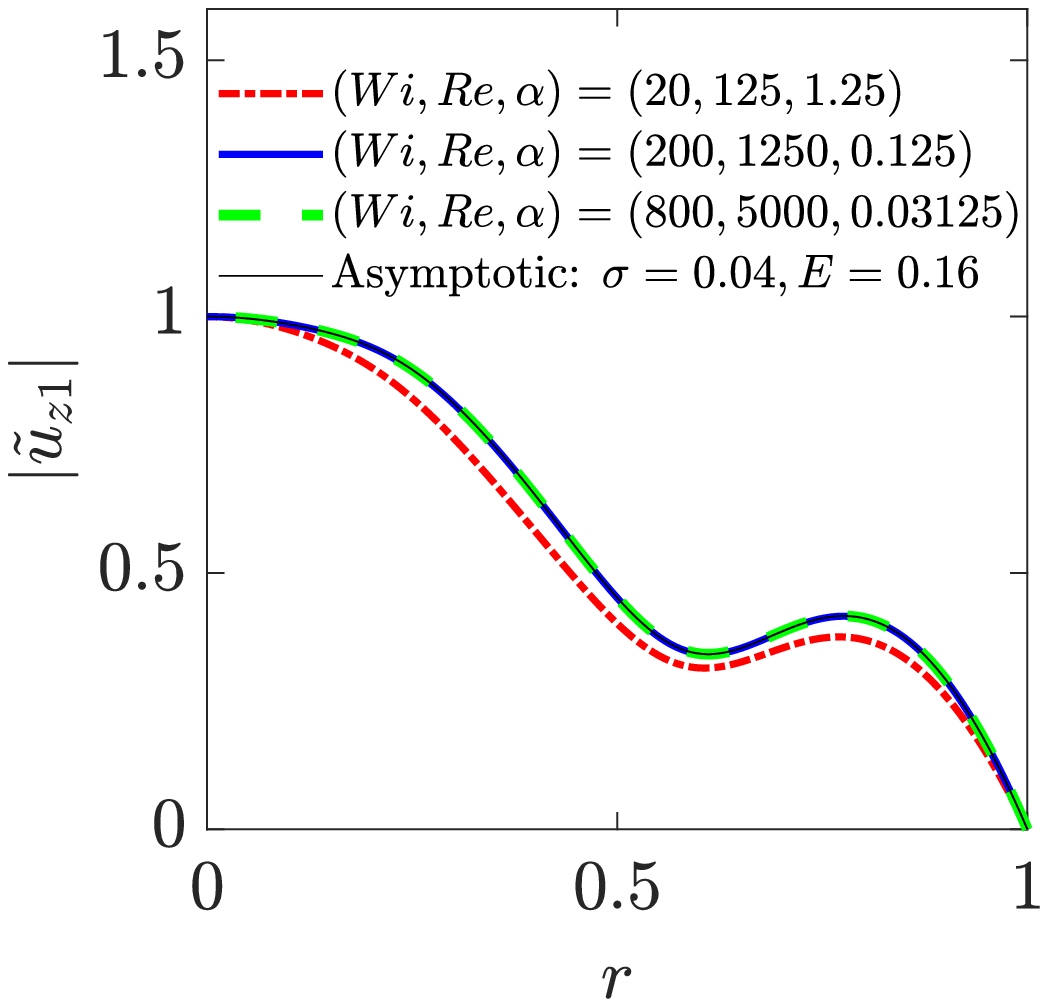}
	\put(-125,107){$(b)$}	
	\caption{$(a)$ Comparison of the eigenspectra obtained from the linear asymptotic equations \eqref{eq:spec-epsilon1} (at $\beta=0.5$, $\sigma=0.04$ and $E=0.16$) and the linear original equation \eqref{eq:original_compact_linear_eigenvalue} (at $\beta=0.5$, $Wi=800$, $Re=5000$ and $\alpha=0.03125$). The eigenvalue $c_1$ in the asymptotic equations is converted to the complex frequency $\omega= \alpha (1+ \sigma c_1)$ for the comparison. $(b)$ The eigenfunction $\tilde u_{z1}$ {(pertaining to the unstable mode)} of the linearised original equations approaches that of the linear asymptotic equations with increasing $Wi$ at $\beta=0.5$.}
	\label{Fig:eigenspectra-eigenfunction}
\end{figure}

{The present calculation is validated by comparing with the results obtained from the original equation system as follows. }
Figure \ref{Fig:eigenspectra-eigenfunction}$(a)$ shows a favourable agreement between the eigenspectrum obtained from the linear asymptotic equations \eqref{eq:spec-epsilon1} for the viscoelastic pipe flow at $\beta=0.5$, $\sigma=0.04$, $E=0.16$ and that obtained from the linear original equation \eqref{eq:original_compact_linear_eigenvalue} with $\beta=0.5$ and a large $Wi=800$ (so $Re=Wi/E=5000$ and $\alpha=1/(\sigma Wi)=0.03125$); the inset highlights the unstable mode.
The eigenfunction $\tilde u_{z1}$ (corresponding to the unstable mode) at $\beta=0.5$, $\sigma=0.04$ and $E=0.16$ is plotted in figure \ref{Fig:eigenspectra-eigenfunction}$(b)$. The eigenfunction in the original equations approaches that of the linear asymptotic solutions as $Wi$ increases, confirming the accuracy of the asymptotic prediction.

\end{appendix}

\bibliographystyle{jfm}
%\bibliography{/Users/mengqizhang/Documents/1_Research/1000_References/BibRef}
\bibliography{references.bib}

\end{document}